\documentclass[aps,amsmath,amssymb,reprint,prd]{revtex4-1}
\usepackage{graphicx}
\usepackage{dcolumn}
\usepackage{bm}

\def \vect#1{\mathbf{#1}}
\def \refeq#1{(\ref{#1})}
\def \refsec#1{Sec.~\ref{#1}}
\def \reffig#1{Fig.~\ref{#1}}
\def \abs#1{\left\vert{#1}\right\vert}
\def \bdij#1{\dot\beta_{#1}}
\def \bij#1{\beta_{#1}}
\def \disc#1{\left.{#1}\right\vert_{\rm disc}}
\def \cont#1{\left.{#1}\right\vert_{\rm cont}}
\def \ii{{\mathsf{i}\,}}
\def \dft#1{\widetilde{#1}}
\def \keff{k^{\rm eff}}
\def \kstd{k^{\rm std}}
\def \vkeff{\vect{k}^{\rm eff}}
\def \vkstd{\vect{k}^{\rm std}}
\def \opr#1{\mathbf{#1}}
\def \mtt#1{{M^{\rm TT}\left({#1}\right)}}
\def \midtilde{{\raise.17ex\hbox{$\scriptstyle\mathtt{\sim}$}}}
\def \figurewidth{0.9\linewidth}
\def \halffigurewidth{0.42\linewidth}
\def \sumlat{\sum_{\rm lattice}}

\begin{document}

\preprint{AIP/123-QED}

\title{The Art of Lattice and Gravity Waves from Preheating \\
{\bf draft for HLattice~V2.0, copyright reserved}}
\author{Zhiqi Huang}
\affiliation{${}^1$ CEA, Institut de Physique Th{\'e}orique, 91191 Gif-sur-Yvette c{\'e}dex, France}

\date{\today}

\begin{abstract}
The nonlinear dynamics of preheating after early-Universe inflation is often studied with lattice simulations. In this work I present a new lattice code HLATTICE. It differs from previous public available codes in the following three aspects: (i) A much higher accuracy is achieved with a modified sixth-order symplectic integrator; (ii) scalar, vector, and tensor metric perturbations in synchronous gauge and their feedback to the dynamics of scalar fields are all included; (iii) the code uses a projector that completely removes the scalar and vector components defined by the discrete spatial derivatives. Such a generic code can have wide range of applications. As an example, gravity waves from preheating after inflation are calculated with a better accuracy.
\end{abstract}

\maketitle

\section{\label{sec:intro} Introduction}
Early-Universe inflation \cite{Guth1981,Guth1982,Starobinsky1982} has become one of the key elements of the standard cosmological model \cite{book_Peacock, book_Mukhanov, book_Weinberg}. In this paradigm the Universe went through a phase of accelerated expansion, which, in the simplest scenario is, driven by a scalar field, namely the inflaton. The predictions of inflation have been confirmed by the high-precision measurements of CMB \cite{Komatsu2010, Fowler2010, Chiang2009, Castro2009, Reichardt2008, Goldstein2003, Sievers2007, Jones2006, Dickinson2004,Hanany2000,Fixsen1996} and the large scale structure surveys \cite{ Reid2009,Eisenstein2005,Cole2005}. Combining these observations with supernova \cite{Astier2006, Miknaitis2007, Wood-Vasey2007,Kessler2009, Garnavich1998, Knop2003,Riess2004,Riess2007}, weak gravitational lensing \cite{Benjamin2007,Massey2007,Lesgourgues2007,Hoekstra2006,Schimd2007,Hoekstra2002a,Hoekstra2002b,Hoekstra2002a,Hoekstra2002b,Van-Waerbeke2005,Schimd2007} and Lyman-$\alpha$ forest data \cite{Viel2004,Kim2004,Croft2002,McDonald2005, McDonald2006}, we find that the current Universe is undergoing another cosmic acceleration \cite{Riess1998, Perlmutter1999,  Bond1999}. This could be due to a cosmological constant \cite{Einstein1917} or another scalar field \cite{Ratra1988,Wetterich1988,Frieman1995,Binetruy1999,Barreiro1999,Brax1999,Copeland2000,Macorra2001,Linder2006,Huterer2007,Linder2008c,Huang2011}. 

It is hence important to understand the dynamics of scalar field(s) in a perturbed Friedmann-Robertson-Walker (FRW) background \cite{book_Peacock,book_Mukhanov, book_Weinberg}. In many cases these scalar fields are almost homogeneous. Thus linear or second-order perturbation theory is enough to describe such a system. However, there are exceptions. Something that we cannot avoid in any successful inflationary model is the decay of the inflaton after inflation. This could start with preheating, a nonperturbative violent process due to parametric resonance or tachyonic growth of fluctuations \cite{Kofman1994,Kofman1997,Greene1997, Greene1998, Felder2001b, Felder2001c, Gumrukcuoglu2008,Barnaby2009c, Braden2010,Frolov2010, Allahverdi2010,Brax2010}. The typical comoving scale of preheating is much smaller than current cosmological scales. However, cosmological-scale comoving curvature fluctuations can also be generated via, e.g., preheating modulated by a field that is light during inflation but becomes heavy at the end of inflation~\cite{Chambers2008,Chambers2008b,Bond2009}. Moreover, the stochastic background of gravity waves (GWs) generated during preheating \cite{Khlebnikov1997, Dufaux2007,Dufaux2009,Dufaux2009b,Dufaux2010,Garcia2007,Garcia2008,Easther2007, Easther2008} is, in principle, observable. In particular, GWs from preheating after hybrid inflation may be observable with the next generation of GW probes \cite{Garcia2008}, although this depends on the parameters in this model. See also \cite{Dufaux2009}, where the parameter space is systematically explored. For more models suggesting potential observables from preheating, the reader is referred to Refs.~\cite{Kasuya1998,Diaz-Gil2008,Tranberg2003, Tranberg2006,Fenu2009,Garcia2011}.

To make quantitative predictions of the observables from preheating, one often needs to run full nonlinear lattice simulations. In the previously mentioned GW calculations, the evolution of scalar fields is done in configuration space using the public available code LATTICEEASY \cite{Felder2008} or unreleased codes with similar techniques. The linear metric perturbations are evolved either in configuration space \cite{Dufaux2009, Dufaux2009b, Dufaux2010, Garcia2007, Garcia2008} or in Fourier space \cite{Dufaux2007, Easther2007,Easther2008}. In these treatments, a traceless-transverse (TT) ``tensor mode'' is defined in Fourier space \cite{book_Weinberg1972}:
\begin{equation}
  h_{ij,\vect{k}}^{\rm TT} = \left[\mtt{\vect{k}}\right]_{ij,lm}h_{lm,\vect{k}}\ , \label{TTdef_Fourier}
\end{equation}
where $h_{ij}$ are the metric perturbations in synchronous gauge \cite{book_Peacock,Ma1995}. The Fourier-space matrix form of the TT projector, $\mtt{\vect{k}}$, is given by
\begin{equation}
  \left[\mtt{\vect{k}}\right]_{ij,lm}\equiv P_{il}\left(\vect{k}\right)P_{mj}\left(\vect{k}\right) -  \frac{1}{2} P_{ij}\left(\vect{k}\right) P_{lm}\left(\vect{k}\right)\ , 
\end{equation}
where
\begin{equation}
  P_{ij}\left(\vect{k}\right) \equiv \delta_{ij} - \frac{k_i k_j}{k^2} \ .\label{Pijdef}
\end{equation}
Here $\delta_{ij}$ is the Kronecker delta. If not otherwise stated, repeated indices are implicitly summed over. The Latin indices run from 1 to 3 (spatial indices). The Greek indices run from 0 to 3 (temporal and spatial indices).

One could also define the TT component in configuration space:
\begin{equation}
  h_{ij} =  \frac{h}{3}\delta_{ij} + \left(\partial_i \partial_j -\frac{1}{3} \delta_{ij}\nabla^2\right) \Lambda + \partial_iA_j + \partial_j A_i + h_{ij}^{\rm TT} \ , \label{TTdef_real}
\end{equation}
where 
\begin{eqnarray}
\nabla^2 \equiv \partial_1^2+\partial_2^2+\partial_3^2 \, \\ 
h \equiv h_{ii}\,,\ \partial_i A_i = 0, \,\partial_i h_{ij}^{\rm TT} = 0\ . \label{gaugecond}
\end{eqnarray}
Here $h_{ij}$ is decomposed into two scalar components ($h$ and $\Lambda$), one vector component ($A_i$) and one tensor component ($h_{ij}^{\rm TT}$). The tracelessness of $h_{ij}^{\rm TT}$ can be confirmed by taking the trace of Eq.~\refeq{TTdef_real} and by using Eq.~\refeq{gaugecond}.

In the continuous case,  definition~\refeq{TTdef_real} is equivalent to definition~\refeq{TTdef_Fourier}. But this is not so for a periodic and cubical box with length $L$ and $n^3$ grid points, in which the spatial derivatives in Eq.~\refeq{TTdef_real} are replaced by finite difference. This is discussed in detail in \refsec{subsec:discretization}. The discrepancy between definitions~\refeq{TTdef_Fourier} and~\refeq{TTdef_real} leads to scalar-tensor mixing: the scalar part of the energy-momentum tensor, calculated in configuration space using discrete derivatives (finite difference), can produce GW in Fourier space defined by Eq.~\refeq{TTdef_Fourier}. At scales where the scalar components dominate, it is difficult to suppress this ``noise'' or distinguish it from the physical GW. Simulations with very high resolution ($n\gtrsim 10^4$), which might solve the problem, are not favored, practically, as they are numerically expensive and often limited by the machine memory. In the new code HLATTICE that I will present in the paper, I define, evolve, and extract scalar, vector, and tensor modes consistently in configuration space. Even though the discrepancy between real (continuous) physics and the numerical (discrete) model cannot be completely removed in any numerical calculations, in HLATTICE the scalar, vector and tensor parts of the metric perturbations are only sourced, respectively, by the scalar, vector, and tensor parts of the energy-momentum tensor. 

In other public available lattice codes, such as LATTICEEASY, DEFROST \cite{Frolov2008}, and CUDAEasy \cite{Sainio2010}, the scalar fields are evolved in a FRW background, and metric perturbations are ignored. But at linear level one can approximately take the energy-momentum tensor of the scalar fields as a source, and evolve GW outside the simulation. (Therefore, the scalar-tensor mixing effect is a problem in post-processing. It should not be regarded as a problem of these lattice codes.) Since this is a linear treatment, the TT component separation could be done at the end of the calculation~\cite{Garcia2008}. Except for Ref.~\cite{Bastero2010}, which I will discuss separately in \refsec{sec:conclusion}, the previous works on GWs from preheating are all based on this (or a similar) approach. In these calculations, the feedback of metric perturbations to the dynamics of scalar fields is ignored or partially ignored. HLATTICE is the first code released that consistently evolve all components of metric perturbations together with the scalar fields. Using HLATTICE I find the metric feedback, as conjectured in previous works, is indeed not a dominating effect, at least for the models studied in this paper.

In some situations, in order to capture some small effects \cite{Bond2009} or energy-insensitive modes \cite{Barnaby2009a}, we need to evolve the system of scalar fields and metric perturbations accurately. In HLATTICE I use a sixth-order symplectic integrator for global evolution, and a fourth-order Runge-Kutta integrator \cite{book_NR} with refined time steps for the noncanonical terms only. The advantage of doing so is that no extra memory cost is required. With this integrator the fractional energy noise of the system can be suppressed to $\lesssim 10^{-12}$. This enables us to check the conservation of the total Hamiltonian, including the tiny contribution from energy carried by gravity waves. This is the first time that we can use the constraint equation to accurately check the numerical accuracy in calculations of GW from preheating.

This paper is organized as follows. In \refsec{sec:HLATTICE} I introduce the HLATTICE code; in \refsec{sec:GW} I use HLATTICE to calculate GW from preheating. I discuss and conclude in \refsec{sec:conclusion}.

\section{HLATTICE code \label{sec:HLATTICE}}
\subsection{Theory}
The system that we consider contains $m$ canonical scalar fields $\phi_1,\phi_2, ..., \phi_m$ with a potential $V\left(\phi_1,\phi_2,...,\phi_m\right)$. The action reads
\begin{equation}
  \mathcal{S} = \int\sqrt{g}\,d^4x \left(\frac{1}{2}\partial^\mu \phi_\ell \partial_\mu \phi_\ell -V + \frac{M_p^2}{2} R \right)\ , \label{action_1}
\end{equation}
where $g \equiv \abs{\det{g_{\mu\nu}}}$, $g_{\mu\nu}$ being the spacetime metric; $M_p$ is the reduced Planck mass, related to Newton's gravitational constant $G_N$ via $M_p \equiv 1/\sqrt{8\pi G_N}$; $R$ is the Ricci scalar \cite{book_Peacock}.

The spacetime metric can be written in synchronous gauge \cite{book_Peacock,book_Mukhanov, book_Weinberg}:
\begin{equation}
  ds^2 = dt^2 - g_{ij}dx^idx^j\ . \label{syn}
\end{equation}
Natural units $c=\hbar = 1$ are used.

In the context of inflation, we are interested in an expanding and spatially flat Universe, where the metric $g_{ij}$ is often written as
\begin{equation}
  g_{ij} = a(t)^2\left(\delta_{ij} + h_{ij}\right) \ . \label{trad}
\end{equation}
In a finite volume $L^3$, we choose $a(t)$ to be the ``scale factor,'' given by
\begin{equation}
  a(t) \equiv \left(\frac{1}{L^3}\int \sqrt{g} d^3x\right)^{1/3} \ . \label{atdef}
\end{equation}
The Hubble parameter is defined as
\begin{equation}
H\equiv\frac{\dot a}{a}\ , \label{Hdef}
\end{equation}
where a dot denotes the derivative with respect to time coordinate $t$.

We will consider weak gravitational fields that satisfy the condition $|h_{ij}|\ll 1$. For a problem that requires long-time integration, a growing nonphysical gauge mode might spoil this condition. This specific case is discussed in \refsec{subsec:gauge}.

In HLATTICE I choose to evolve the following variables
\begin{equation}
  \bij{ij} \equiv \left(\ln \mathcal{G}\right)_{ij}\ ,
\end{equation}
where $\mathcal{G}$ is the $3\times 3$ matrix $g_{ij}$. The matrix function $\ln \mathcal{G}$ should be understood as 
\begin{equation}
  \ln \mathcal{G}\equiv 2\mathcal{E}\ln a  + (a^{-2}\mathcal{G}-\mathcal{E}) - \frac{1}{2} (a^{-2}\mathcal{G}-\mathcal{E})^2 + \frac{1}{3} (a^{-2}\mathcal{G}-\mathcal{E})^3 - ... \ ,
\end{equation} 
where $\mathcal{E}$ is the $3\times 3$ identity matrix. To the linear order, we have
\begin{equation}
h_{ij} \approx \beta_{ij} - 2\delta_{ij}\ln a\ \label{hijbetaij}.
\end{equation}

It follows from Eq.~\refeq{hijbetaij} that the traceless part of $\beta_{ij}$, which we define as $\gamma_{ij}$, satisfies
\begin{equation}
\gamma_{ij} \equiv \beta_{ij} - \frac{\beta}{3}\delta_{ij} = h_{ij} - \frac{h}{3}\delta_{ij} + O(h_{ij}^2)\ . \label{bijsmall}
\end{equation}
This implies that $O(\gamma_{ij}^n) \lesssim O(h_{ij}^n)$ for arbitrary $n\ge 0$. This significantly simplifies the procedure to expand an arbitrary function of $\beta_{ij}$ to a given order in $h_{ij}$-- we just need to replace $\beta_{ij}$ with ${\beta \over 3}\delta_{ij} +\gamma_{ij}$ and cut the Taylor expansions in $\gamma_{ij}$ at the same order. In the rest of this subsection, we will do this for the metric and for the action.

Let us first calculate the volume weight $\sqrt{g}$. Using the fact that $\det (\mathcal{G}) \equiv e^{{\rm Tr}(\ln \mathcal{G})}$, we find a simple and exact expression:
\begin{equation}
\sqrt{g} = e^{\beta/2}\ , \label{sqrtg}
\end{equation}
where  $\beta \equiv \beta_{ii}$ is the trace of $\beta_{ij}$.

Writing the $3\times 3$ matrix $\gamma_{ij}$ as $\Gamma$, we can rewrite the $3\times 3$ matrix $g_{ij}$ as
\begin{equation}
\mathcal{G} = e^{{\beta \over 3}\mathcal{E}+\Gamma} = e^{\beta/3}\left(1+\Gamma + \frac{1}{2}\Gamma^2 + {1\over 6} \Gamma^3+ ...\right)\ .
\end{equation}

By taking the inverse of $\mathcal{G}$ we obtain 
\begin{equation}
g^{ij} = e^{-{\beta \over 3}}\left(\delta_{ij}-\gamma_{ij} + \frac{1}{2} \gamma_{ik}\gamma_{kj}\right) + O\left(h_{ij}^3\right).
\end{equation}

We can substitute, for example, $1-\gamma_{11} + \frac{1}{2}\gamma_{11}^2$ with $e^{-\gamma_{11}} + O(h_{ij}^3)$. With such substitutions and ignoring $O(h^3_{ij})$ terms, we can write $g^{ij}$ as functions of $\beta_{ij}$:
\begin{eqnarray}
g^{11} &\approx& e^{-\bij{11}} + \frac{e^{-2\bij{11}/3}}{2} \left(\bij{12}^2e^{-\bij{22}/3} + \bij{13}^2e^{-\bij{33}/3}\right), \nonumber \\
g^{22} &\approx& e^{-\bij{22}} + \frac{e^{-2\bij{22}/3}}{2}  \left(\bij{23}^2e^{-\bij{33}/3} + \bij{21}^2e^{-\bij{11}/3}\right), \nonumber \\
g^{33} &\approx& e^{-\bij{33}} + \frac{e^{-2\bij{33}/3}}{2}  \left(\bij{31}^2e^{-\bij{11}/3} + \bij{32}^2e^{-\bij{22}/3}\right), \nonumber \\
g^{23}&\approx& -\bij{23} e^{-(\bij{22}+\bij{33})/2}+{1\over 2} \bij{12}\bij{31} e^{-\beta/3}\ ,\nonumber \\
g^{31}&\approx& -\bij{31} e^{-(\bij{33}+\bij{11})/2}+{1\over 2} \bij{23}\bij{12} e^{-\beta/3}\ ,\nonumber \\
g^{12}&\approx& -\bij{12} e^{-(\bij{11}+\bij{22})/2}+{1\over 2} \bij{31}\bij{23} e^{-\beta/3}\ . \label{gijeq}
\end{eqnarray}
The difference between the left-hand side and the right-hand side of each equation is less than or equal to $O(h_{ij}^3)$. It is obvious that such approximations are not unique, as one can add arbitrarily more higher-order terms on the right-hand side of each equation. The specific choice I made in Eqs.~\refeq{gijeq} is based on three criteria: 
\begin{itemize}
\item{simplicity;}
\item{less $O(h_{ij}^3)$ residual error terms;}
\item{under a coordinate transformation $x^i \rightarrow C x^i$ ($i=1,\ 2,\ 3$, $C$ is a constant; the metric transforms as $\beta_{ij}\rightarrow \beta_{ij}-2\delta_{ij}\ln C$), the approximated expression of $g^{ij}$ scales as $C^{2}$, just as the exact $g^{ij}$ should.}
\end{itemize}
The last requirement enforces the gradient energy density $g^{ij}\partial_i\phi_\ell\partial_j\phi_\ell$, with the approximations of $g^{ij}$ in Eqs.~\refeq{gijeq} being used, to remain exactly invariant. This allows us to perform a spatial coordinate transformation $x^i\rightarrow Cx^i$ without producing any extra error terms. This can be used to optimize the HLATTICE code. In HLATTICE, after every few evolution steps the coordinate redefinition $x^i\rightarrow Cx^i$ is performed, where $C$ is chosen to be the scale factor {\it at the moment}. After such a transformation, the scale factor is redefined to be $1$ and $\beta_{ij}$ is made small. The exponential functions in Eqs.~\refeq{gijeq} can thus be evaluated via the approximation
\begin{equation}
e^x \approx \left\{\left[\left(\left(\sum_{s=0}^6 \frac{(x/16)^s}{s!}\right)^2\right)^2\right]^2\right\}^2\ ,\ {\rm for}\ |x|\ll 1. \label{exapp}
\end{equation}
For a program built with Fortran code, the evaluation of the right-hand side of \refeq{exapp} is about twice as fast as that of $e^x$. This significantly improves the performance of HLATTICE. 

Confusion should be avoided here. The FRW background is intrinsically different from a Minkowski background.  In the calculation, by adaptively changing the coordinate system, it is possible to keep the scale factor close to $1$ and to keep $\beta_{ij}$ small. However, when deriving the theoretical formulas, we should work in a fixed coordinate system and {\it cannot} assume that $\beta_{ij}$ is small.

The action~\refeq{action_1} can be written as
\begin{equation}
S= \int dt\, \left( K_f - G_f - V_f + K_g - G_g\right)  \ , \label{action_2}
\end{equation}
where $K_f$ is the {\it kinetic energy of the scalar fields},
\begin{equation}
K_f = \int e^{\beta/2}d^3x\, \frac{1}{2} \dot\phi_\ell^2 \ ; \label{kfdef}
\end{equation}
$G_f$ is the {\it gradient energy of the scalar fields},
\begin{equation}
G_f = \int e^{\beta/2}d^3x\, \frac{1}{2} g^{ij} \partial_i \phi_\ell\partial_j \phi_\ell \ ; \label{gfdef}
\end{equation}
$V_f$ is the {\it potential energy of the scalar fields},
\begin{equation}
V_f = \int e^{\beta/2}d^3x\, V(\phi_1,\phi_2,...,\phi_m)\ ; \label{vfdef}
\end{equation}
$K_g$ is the {\it ``kinetic energy'' of gravity}, approximated to second order in $h_{ij}$ by
\begin{eqnarray}
K_g &\approx& \frac{M_p^2}{4} \int e^{\beta/2}d^3x \nonumber\\
&&\times \left( \bdij{23}^2+ \bdij{31} ^2 + \bdij{12} ^2 \right.\nonumber \\ &&\left. - \bdij{11}\bdij{22} - \bdij{22}\bdij{33} - \bdij{33}\bdij{11}\right) \ ; \label{kgdef}
\end{eqnarray}
and $G_g$ is the {\it ``gradient energy'' of gravity}, approximated to second order in $h_{ij}$ by
\begin{eqnarray}
G_g &\approx& \frac{M_p^2}{4} a(t) \int d^3x\,    \nonumber \\
&& \times \left( \bij{23,1}^2 + \bij{31,2}^2 + \bij{12,3}^2 \right. \nonumber  \\ 
&& - 2 \bij{23,1} \bij{31,2}  - 2 \bij{31,2} \bij{12,3} - 2\bij{12,3}\bij{23,1}\nonumber \\
&& - \bij{22,1} \bij{33,1} - \bij{33,2} \bij{11,2}-  \bij{11,3}\bij{22,3} \nonumber \\
&& \left.  + 2\bij{23,2} \bij{11,3} + 2 \bij{31,3}\bij{22,1} + 2 \bij{12,1}\bij{33,2}\right) \ , \label{ggdef}
\end{eqnarray}
where $\beta_{ij,k}\equiv \partial_k\bij{ij}$. In Eq.~\refeq{ggdef} I have approximated the local volume weight $e^{\beta/6}$ with a global scale factor $a(t)$. This is valid since the integrand in Eq.~\refeq{ggdef} is of second order in $h_{ij}$. In Eq.~\refeq{kgdef} such a replacement is not allowed, as the integrand is of zeroth order. With the bilinear approximation \refeq{ggdef}, HLATTICE cannot capture the gravity self-interaction, which in principle can be important on small scales. This is discussed in \refsec{sec:conclusion}.

Equations~(\ref{kgdef}-\ref{ggdef}) are obtained using the same technique that I used to derive Eqs.~\refeq{gijeq}, i.e., rewriting $\beta_{ij}$ as $\gamma_{ij}+{\beta \over 3} \delta_{ij}$ and cutting the Taylor expansions of $\gamma_{ij}$ at the second order. The simplicity of the final expression of $K_g$ is the main reason why I have used $\beta_{ij}$ as fundamental variables in HLATTICE.

It is also easy to verify that the right-hand sides of Eqs.~(\ref{kfdef}-\ref{ggdef}) are all exactly invariant under the spatial-coordinate redefinition $x^i\rightarrow Cx^i$.

\subsection{The discretization scheme \label{subsec:discretization}}
In this subsection I answer or attempt to answer the following questions:
\begin{itemize}
\item[1.]{What exactly is calculated in a lattice simulation?}
\item[2.]{How do we choose a discretization scheme to include metric perturbations?}
\item[3.]{How do we define the scalar, vector, and tensor on the lattice?}
\end{itemize}

Before answering these questions, let us briefly review the lattice theory and the discrete Fourier transformation (DFT).

I label a grid point in the lattice with three integer numbers $(i_1, i_2, i_3)$ and apply the periodic boundary condition (PBC) 
\begin{equation}
f_{i_1+n,i_2,i_3}=f_{i_1,i_2+n,i_3} = f_{i_1, i_2, i_3+n} = f_{i_1,i_2,i_3}\ , \label{pcond}
\end{equation}
where $n$ is a fixed integer representing the resolution of the simulation; $f$ represents all physical quantities including the scalar fields, the metric, and their temporal/spatial derivatives. Given the PBC, we only need to evolve $f_{i_1,i_2,i_3}$ in a cubical fundamental box containing $n^3$ grid points ($-n/2<i_1,i_2,i_3 \le n/2$). In what follows I will use the notation $\sumlat$ to denote the summation over grid points in this fundamental box.

The DFT is defined as \cite{book_NR}
\begin{equation}
\dft{f}_{j_1, j_2, j_3} \equiv \sumlat e^{-\frac{2\pi\ii}{n}( i_1j_1 + i_2j_2 + i_3j_3)}f_{i_1,i_2,i_3} \ , \label{dftdef}
\end{equation}
where $\ii$ is the imaginary unit and $j_1, j_2, j_3$ are arbitrary integers. Unless otherwise stated, the overhead tilde sign represents variables in Fourier space. 

It is easy to verify that in Fourier space $\dft{f}_{j_1, j_2, j_3}$  also satisfies the PBC
\begin{equation}
\dft{f}_{j_1+n, j_2, j_3}=\dft{f}_{j_1, j_2+n, j_3}=\dft{f}_{j_1, j_2, j_3+n}=\dft{f}_{j_1, j_2, j_3} \ .
\end{equation}
Moreover, if a field satisfies the PBC in Fourier space, it also satisfies the PBC in configuration space. Thus in what follows we no longer distinguish between the two PBCs. Similarly we define a Fourier-space fundamental box with $-n/2<j_1,j_2,j_3 \le n/2$.  

The {\it standard DFT wave vector} is defined as 
\begin{equation}
\vkstd_{j_1,j_2,j_3} \equiv \frac{2\pi}{L}(j_1,\, j_2,\, j_3)\ , \label{stdk}
\end{equation}
where $L$ is the coordinate length of each edge of the fundamental box in configuration space. The amplitude of $\vkstd_{j_1,j_2,j_3} $ is 
\begin{equation}
\kstd_{j_1,j_2,j_3} =\frac{2\pi}{L}\sqrt{j_1^2+j_2^2+j_3^2}\ .
\end{equation}

I will use the subscripts ``disc'' and ``cont'' to distinguish between quantities evaluated on the discrete lattice or in a continuum. These subscripts are often omitted when it does not give rise to ambiguity.

\subsubsection{What exactly is calculated in a lattice simulation?}
In LATTICEEASY, DEFROST, and CUDAEasy, the metric perturbations are ignored. Scalar fields are evolved in configuration space. The lattice-version equation of motion (EOM) of the scalar fields is 
\begin{equation}
\left(\frac{d^2}{dt^2} -\frac{\nabla^2}{a^2} + 3H\frac{d}{dt}\right)\left.\phi_\ell\right\vert_{i_1,i_2,i_3} + \left.\frac{\partial V}{\partial\phi_\ell}\right\vert_{i_1,i_2,i_3} = 0 \ , \label{diseom}
\end{equation}
where $\nabla^2$ is the discrete Laplacian operator. 

In this subsection we will discuss an idealized case where the temporal differential operator $d/dt$ can be perfectly integrated. Moreover, the numerical errors in $a(t)$ and other background quantities are assumed to be negligible. These assumptions can be quite close to the reality, if a high-order integrator is used in the lattice code.

The only spatial operator that appears in the EOM is the Laplacian operator $\nabla^2$, which can be directly defined without referring to any first-order discrete derivatives. For example, in LATTICEEASY $\nabla^2$ is defined as
\begin{eqnarray}
\left.\nabla^2 f\right\vert_{i_1,i_2,i_3} &\equiv& \frac{1}{\Delta^2}\left(f_{i_1+1,j_1,k_1}+f_{i_1-1,j_1,k_1}\right. \label{LEnabla2} \\
&& +f_{i_1,j_1+1,k_1} +f_{i_1,j_1-1,k_1} \nonumber \\
&& \left.+f_{i_1,j_1,k_1+1}+f_{i_1,j_1,k_1-1}-6f_{i_1,j_1,k_1}\right)\nonumber\ ,
\end{eqnarray}
where $\Delta \equiv L/n$ is the coordinate distance between neighboring points in configuration space. By taking the DFT of the above equation, we obtain its equivalent form in Fourier space,
\begin{equation}
\left. \dft{\nabla^2 f}\right\vert_{j_1,j_2,j_3} = -\left(\keff_{j_1,j_2,j_3}\right) ^2 \dft{f}_{j_1,j_2,j_3}, \label{LEnabla2fdef}
\end{equation}
where
\begin{equation}
\left(\keff_{j_1,j_2,j_3}\right) ^2\equiv \frac{4}{\Delta^2}\left(\sin^2{\frac{\pi j_1}{n}} +\sin^2{\frac{\pi j_2}{n}}+\sin^2{\frac{\pi j_3}{n}}\right). \label{LEkeff2}
\end{equation}
Because both $\dft{\nabla^2f}$ and $\dft{f}$ satisfy the PBC, their ratio $-(\keff)^2$ should also. This can be explicitly checked in Eq.~\refeq{LEkeff2}.

The reader should keep in mind that Eq.~\refeq{LEkeff2} is a consequence of the LATTICEEASY definition of discrete $\nabla^2$. In other lattice codes the definition of discrete $\nabla^2$ (and hence $\keff$) can be different.

Using Eq.~\refeq{LEnabla2fdef} we rewrite the EOM, Eq.~\refeq{diseom} in Fourier space:
\begin{eqnarray}
 \left[\frac{d^2}{dt^2} +  \left(\frac{\keff_{j_1,j_2,j_3}}{a}\right) ^2  + 3H\frac{d}{dt}\right]\left.\dft{\phi}_\ell\right\vert_{j_1,j_2,j_3} \nonumber \\ 
 + \left.\dft{\left(\frac{\partial V}{\partial \phi_\ell}\right)}\right\vert_{j_1,j_2,j_3} = 0\ . \label{diseomdft}
\end{eqnarray}

Equation~\refeq{diseomdft} is the equation being integrated on the lattice. Comparing it to the continuous-case EOM for a mode $k=\keff_{j_1,j_2,j_3}$, we find that the only discrepancy comes from the difference between the DFT of $\partial V/\partial \phi_\ell$ and the continuous Fourier transformation of it. This discrepancy is model dependent and difficult to quantify in the nonlinear regime. However, in the linear regime, as I will show below, this discrepancy vanishes.

In the linear regime, we can rewrite the lattice-version EOM in configuration space as
\begin{eqnarray}
\left(\frac{d^2}{dt^2} -\frac{\nabla^2}{a^2} + 3H\frac{d}{dt}\right)\left.\delta\phi_\ell\right\vert_{i_1,i_2,i_3} \nonumber \\
 +\langle\frac{\partial^2 V}{\partial\phi_{\ell'}\partial\phi_\ell}\rangle\left.\delta\phi_{\ell'}\right\vert_{i_1,i_2,i_3}  = 0 \ . \label{diseompt}
\end{eqnarray}
where $\langle\cdot\rangle$ represents the lattice average $\frac{1}{n^3}\sumlat\cdot $. The field perturbation $\delta\phi_\ell$ is defined as
\begin{equation}
\left.\delta\phi_\ell\right\vert_{i_1,i_2,i_3} \equiv\left. \phi_\ell\right\vert_{i_1,i_2,i_3} -\langle\phi_\ell\rangle \ .
\end{equation}

The Fourier-space version of Eq.~\refeq{diseompt} is
\begin{eqnarray}
 \left[\frac{d^2}{dt^2} +  \left(\frac{\keff_{j_1,j_2,j_3}}{a}\right) ^2  + 3H\frac{d}{dt}\right]\left.\dft{\delta\phi}_\ell\right\vert_{j_1,j_2,j_3} \nonumber \\
+ \langle\frac{\partial^2 V}{\partial\phi_{\ell'}\partial\phi_\ell}\rangle\left.\dft{\delta\phi}_{\ell'}\right\vert_{j_1,j_2,j_3} = 0\ , \label{diseomptdft}
\end{eqnarray}
which is {\it identical} to the continuous-case EOM for a linear mode with wavenumber $k=\keff$. If we interpret $\keff$ as the physical wavenumber, in the limit that a perfect integrator is used, all the linear modes will be correctly solved on the lattice. In other words, in a lattice code with a good integrator, the difference between a discrete system and a continuum does not matter in the linear regime. 

What has been discussed above is well known to the lattice community. However, in many previous works, $\kstd$ in the fundamental box is interpreted as the physical wavenumber \cite{Dufaux2006, Felder2008, Dufaux2007,Dufaux2009,Dufaux2009b,Dufaux2010,Garcia2007,Garcia2008,Easther2007, Easther2008}. This is usually not a serious problem. With a good definition of discrete $\nabla^2$, $\keff$ should be close to $\kstd$ in the fundamental box. However, there are exceptions where the difference between $\keff$ and $\kstd$ is relevant. One specific case is the calculation of GWs that I will discuss later. 

\subsubsection{How do we choose a discretization scheme to include metric perturbations?} 

We have seen that defining the discrete $\nabla^2$ in configuration space is equivalent to defining a kernel $-(\keff)^2$ in Fourier space. First of all, the kernel should satisfy the PBC. Second, since the linear modes on the lattice behave like linear modes with $k=\keff$ in a continuum, we should restrict $(\keff)^2$ to be positive to avoid nonphysical tachyonic growth. Third, for practical purpose, in configuration space the discrete operator $\nabla^2$ should involve only a few neighbors. Thus the functional form of $(\keff \Delta)^2$ is limited to be a low-order polynomial of $e^{\pm 2\ii\pi j_1/n}$, $e^{\pm 2\ii\pi j_2/n}$, and $e^{\pm 2\ii\pi j_3/n}$. Finally, we want to choose a $(\keff)^2$ that is close to $(\kstd)^2$, as we want the nonlinear-regime mode-mode coupling on the lattice to mimic the real physics in a continuum.

Using Eq.~\refeq{LEnabla2}, the reader can verify that the LATTICEEASY discrete $\nabla^2$ satisfies all the above conditions. Moreover, in configuration space it is accurate to the linear order of $\Delta$:
\begin{equation}
\disc{\nabla^2f} = \cont{\nabla^2f} + O(\Delta^2)\ .
\end{equation}
While this discretization scheme is simple and reasonably accurate, there may be various reasons to improve it. For example, the discretization scheme in DEFROST improves the isotropy of $\nabla^2$ without much additional computational cost \cite{Frolov2008}. In DEFROST the default discrete $\nabla^2$ is defined as 
\begin{equation}
\nabla^2f_{i_1,i_2,i_3} \equiv \sum_{-1\le i_1',i_2',i_3'\le 1}C_{|i_1'|+|i_2'|+|i_3'|}f_{i_1+i_1',i_2+i_2',i_3+i_3'},  \label{DEFROSTnabla2}
\end{equation}
where $C_0=-64/15$, $C_1=7/15$, $C_2= 1/10$, and $C_3=1/30$.

For HLATTICE, which includes metric perturbations, however, the discretization problem is much more complicated, for the reasons that I list below.
\begin{itemize}
\item[1.]{The governing equations for the metric perturbations are rather complicated. It is possible that nonphysical tachyonic growth arises in metric perturbations even when the discrete $\nabla^2$ is negative definite.}
\item[2.]{The field EOMs now involve first-order derivatives of the fields, whose discrete forms should be properly defined and should be consistent with the lattice $\nabla^2$.}
\item[3.]{With the metric perturbations, the gradient energy terms ($G_f$ and $G_g$) are much more complicated. This prevent us from choosing complicated discretization schemes, whose computational cost would be intolerable.}
\end{itemize}
To address the first point, I run HLATTICE for empty spacetime with initial small noises in the metric, and verify that the noises do not grow. I propose that such a null test should be done for any discretization schemes of gravity.

Following the last two points, in the first released version of HLATTICE (HLATTICE~V1.0) I have used the simplest discrete $\nabla$ that is accurate to the linear order of $\Delta$:
\begin{eqnarray}
\left.\partial_1\right\vert_{i_1,i_2,i_3} f &\equiv& \frac{1}{2\Delta}\left(f_{i_1+1, i_2, i_3} - f_{i_1-1,i_2,i_3}\right)\ , \nonumber \\
\left.\partial_2 f\right\vert_{i_1,i_2,i_3} &\equiv&\frac{1}{2\Delta}\left(f_{i_1, i_2+1, i_3} - f_{i_1,i_2-1,i_3}\right)\ , \label{derv1} \\
\left.\partial_3 f\right\vert_{i_1,i_2,i_3} &\equiv& \frac{1}{2\Delta}\left(f_{i_1, i_2, i_3+1} - f_{i_1,i_2,i_3-1}\right) \ . \nonumber
\end{eqnarray}

The equivalent Fourier-space form can be achieved by taking DFT of Eqs.~\refeq{derv1}. The result is:
\begin{eqnarray}
\dft{\partial_1 f}_{j_1,j_2,j_3} = \frac{\ii}{\Delta}\sin \left(\frac{2\pi j_1}{n}\right) \dft{f}_{j_1,j_2,j_3}\ , \nonumber \\
\dft{\partial_2 f}_{j_1,j_2,j_3} = \frac{\ii}{\Delta}\sin \left(\frac{2\pi j_2}{n}\right) \dft{f}_{j_1,j_2,j_3}\ ,  \\
\dft{\partial_3 f}_{j_1,j_2,j_3} = \frac{\ii}{\Delta}\sin \left(\frac{2\pi j_3}{n}\right) \dft{f}_{j_1,j_2,j_3}\ . \nonumber
\end{eqnarray}
The above equations can be written in a more compact form
\begin{equation}
\dft{\nabla f}_{j_1,j_2,j_3} = \ii \vkeff_{j_1,j_2,j_3} \dft{f}_{j_1,j_2,j_3} \ , \label{nablafdftH1}
\end{equation}
where the effective wave vector $\vkeff_{j_1,j_2,j_3}$ is
\begin{equation}
\vkeff_{j_1,j_2,j_3} = \frac{1}{\Delta}\left(\sin{\frac{2\pi j_1}{n}},\, \sin{\frac{2\pi j_2}{n}},\, \sin{\frac{2\pi j_3}{n}}\right)\ . \label{HL1effkdef}
\end{equation}

Thus the Fourier-space kernel for $-\nabla^2$ is \\
\begin{equation}
(\keff_{j_1,j_2,j_3})^2 = \frac{1}{\Delta^2} \left(\sin^2{\frac{2\pi j_1}{n}}+\sin^2{\frac{2\pi j_2}{n}}+\sin^2{\frac{2\pi j_3}{n}}\right)\ . \label{HL1keff2}
\end{equation}
Either taking the inverse DFT of Eq.~\refeq{HL1keff2} or repeatedly using Eq.~\refeq{derv1}, we obtain the configuration-space definition of discrete $\nabla^2$ in HLATTICE~V1.0:
\begin{eqnarray}
\left.\nabla^2 f\right\vert_{i_1,i_2,i_3} &=& \frac{1}{4\Delta^2}\left(f_{i_1+2,j_1,k_1}+f_{i_1-2,j_1,k_1}\right. \label{HL1nabla2} \\
&& +f_{i_1,j_1+2,k_1} +f_{i_1,j_1-2,k_1} \nonumber \\
&& \left.+f_{i_1,j_1,k_1+2}+f_{i_1,j_1,k_1-2}-6f_{i_1,j_1,k_1}\right)\nonumber\ .
\end{eqnarray}
This discrete $\nabla^2$ is also negative definite and accurate to linear order of $\Delta$. However, with this definition of $\nabla^2$, if the metric backreaction is negligible and $n$ is an even number, a grid point on the lattice will never interact with its neighbors. The Fourier-space point of view is that $\keff$ in Eq.~\refeq{HL1keff2} satisfies a stronger PBC:
\begin{equation}
\keff_{j_1+n/2,j_2,j_3}=\keff_{j_1,j_2+n/2,j_3}=\keff_{j_1,j_2,j_3+n/2}=\keff_{j_1,j_2,j_3}\ ,
\end{equation}
which indicate that we are only studying modes within a smaller fundamental box. The higher modes (interaction between neighboring points) are not included in this discretization scheme.

Being unsatisfied with the discretization scheme in HLATTICE~V1.0, I introduced a new discretization scheme in the current version of HLATTICE, HLATTICE~V2.0. The discrete derivatives are defined as
\begin{eqnarray}
\left.\partial_1\right\vert_{i_1,i_2,i_3} f &\equiv& \frac{1}{12\Delta}\left[8\left(f_{i_1+1, i_2, i_3} - f_{i_1-1,i_2,i_3})\right.\right. \nonumber \\
&&\left.\left.-(f_{i_1+2,i_2,i_3}-f_{i_1-2,i_2,i_3}\right)\right]\ , \nonumber \\
\left.\partial_2\right\vert_{i_1,i_2,i_3} f &\equiv& \frac{1}{12\Delta}\left[8\left(f_{i_1, i_2+1, i_3} - f_{i_1,i_2-1,i_3})\right.\right. \label{derv2} \\
&&\left.\left.-(f_{i_1,i_2+2,i_3}-f_{i_1,i_2-2,i_3}\right)\right]\ , \nonumber \\
\left.\partial_3\right\vert_{i_1,i_2,i_3} f &\equiv& \frac{1}{12\Delta}\left[8\left(f_{i_1, i_2, i_3+1} - f_{i_1,i_2,i_3-1})\right.\right. \nonumber \\
&&\left.\left.-(f_{i_1,i_2,i_3+2}-f_{i_1,i_2,i_3-2}\right)\right]\ , \nonumber
\end{eqnarray}
whose Fourier-space form is
\begin{equation}
\dft{\nabla f}_{j_1,j_2,j_3} = \ii \vkeff_{j_1,j_2,j_3} \dft{f}_{j_1,j_2,j_3} \ , \label{nablafdftH2}
\end{equation}
where the effective wave vector $\vkeff_{j_1,j_2,j_3}$ is
\begin{eqnarray}
\vkeff_{j_1,j_2,j_3} = \frac{1}{3\Delta}\left(\sin{\frac{2\pi j_1}{n}}\left(4-\cos{\frac{2\pi j_1}{n}}\right),\right. \nonumber \\
\sin{\frac{2\pi j_2}{n}}\left(4-\cos{\frac{2\pi j_2}{n}}\right),\nonumber \\ 
\left.\sin{\frac{2\pi j_3}{n}}\left(4-\cos{\frac{2\pi j_3}{n}}\right)\right)\ . \label{HL2effkdef}
\end{eqnarray}
This discrete $\nabla$ is two-orders more accurate: $\disc{\nabla f} =\cont{\nabla f} + O(\Delta^4)$.

Explicitly written in configuration space, the discrete $\nabla^2$ in HLATTICE~V2.0 is
\begin{eqnarray}
\left.\nabla^2f\right\vert_{i_1,i_2,i_3} &\equiv& \frac{1}{144} \Bigl[f_{i_1+4,i_2,i_3}+f_{i_1-4,i_2,i_3} +f_{i_1,i_2+4,i_3} \nonumber \\
&& +f_{i_1,i_2-4,i_3}+f_{i_1,i_2,i_3+4} +f_{i_1,i_2,i_3-4} \nonumber \\
&& +16\bigl(f_{i_1+1,i_2,i_3}+f_{i_1-1,i_2,i_3}+f_{i_1,i_2+1,i_3}\nonumber \\
&& +f_{i_1,i_2-1,i_3}+f_{i_1,i_2,i_3+1}+f_{i_1,i_2,i_3-1} \nonumber \\
&& -f_{i_1+3,i_2,i_3}-f_{i_1-3,i_2,i_3}-f_{i_1,i_2+3,i_3}\nonumber \\
&& -f_{i_1,i_2-3,i_3}-f_{i_1,i_2,i_3+3}-f_{i_1,i_2,i_3-3}\bigr)\nonumber \\
&& +64\bigl(f_{i_1+2,i_2,i_3}+f_{i_1-2,i_2,i_3}+f_{i_1,i_2+2,i_3}\nonumber \\
&& +f_{i_1,i_2-2,i_3}+f_{i_1,i_2,i_3+2}+f_{i_1,i_2,i_3-2}\bigr)\nonumber \\
&& -390f_{i_1,i_2,i_3}\Bigr] \label{HL2nabla2}
\end{eqnarray}

Finally, I summarize all the discretization schemes in Table~\ref{tbldisc}.
\begin{table}
\caption{Discretization schemes of lattice codes. The ``accuracy'' is defined as $(\cont{\nabla^2f} -\disc{\nabla^2f})$. \label{tbldisc}}
\begin{tabular}{llll}
\hline
\hline
Code & $\nabla$ & $\nabla^2$ & Accuracy  \\
\hline
LATTICEEASY\ \   & Undefined\ \ & Eq.~\refeq{LEnabla2}\ \  & $O(\Delta^2)$ \\
DEFROST &  Undefined & Eq.~\refeq{DEFROSTnabla2} & $O(\Delta^2)$ \\
CUDAEasy &  Undefined & Eq.~\refeq{DEFROSTnabla2} & $O(\Delta^2)$ \\
HLATTICE~V1.0 & Eqs.~\refeq{derv1} & Eq.~\refeq{HL1nabla2} & $O(\Delta^2)$ \\ 
HLATTICE~V2.0 & Eqs.~\refeq{derv2} & Eq.~\refeq{HL2nabla2} & $O(\Delta^4)$ \\
\hline
\hline
\end{tabular}
\end{table}

\subsubsection{How do we define the scalar, vector, and tensor on the lattice?}
In previous works \cite{Dufaux2006, Felder2008, Dufaux2007,Dufaux2009,Dufaux2009b,Dufaux2010,Garcia2007,Garcia2008,Easther2007, Easther2008}, the scalar, vector and tensor part of $h_{ij}$ are defined in Fourier space:
\begin{eqnarray}
  \dft{h}_{ij} &=&  \frac{\dft{h}}{3}\delta_{ij} - \left(\kstd_i \kstd_j - \frac{1}{3} \delta_{ij} (\kstd)^2\right) \dft{\Lambda}\nonumber  \\
&& + \ii\kstd_i\dft{A}_j + \ii\kstd_j \dft{A}_i + \dft{h}_{ij}^{\rm TT} \ , \label{TTdef_std}
\end{eqnarray}
where 
\begin{equation}
\kstd_i \dft{A}_i = \kstd_i \dft{h}^{\rm TT}_{ij} = \dft{h}^{\rm TT}_{ii}=0\ . \label{stdconstr}
\end{equation}

Because $\kstd$ does not satisfy the PBC, definition~\refeq{TTdef_std} has to be limited in the fundamental box or in a smaller region. It is clear that this definition is not equivalent to the discrete form of Eq.~\refeq{TTdef_real}, whose Fourier-space counter-part is
\begin{eqnarray}
  \dft{h}_{ij} &=&  \frac{\dft{h}}{3}\delta_{ij} - \left(\keff_i \keff_j - \frac{1}{3} \delta_{ij} (\keff)^2\right) \dft{\Lambda}\nonumber  \\
&& + \ii\keff_i\dft{A}_j + \ii\keff_j \dft{A}_i + \dft{h}_{ij}^{\rm TT} \ , \label{TTdef_eff}
\end{eqnarray}
where 
\begin{equation}
\keff_i \dft{A}_i = \keff_i \dft{h}^{\rm TT}_{ij} = \dft{h}^{\rm TT}_{ii}=0\ .
\end{equation}

Since the Fourier modes of scalar fields on the lattice follow EOM~\refeq{diseomdft}, Eq.~\refeq{TTdef_eff} seems to be a more proper definition. Thus, the question is whether one should use $\mtt{\vkstd}$ or $\mtt{\vkeff}$ as the TT projector. One may argue that with a reasonable UV cutoff, the difference between $\vkeff$ and $\vkstd$ is small. However, the TT projection is a subtle procedure. Because the scalar metric perturbations are typically much larger than the tensor ones, we are extracting a small number from a big number. A little difference in the projector may lead to a big error. With $n\sim 10^2$-$10^3$ and a simple discretization scheme, most of the modes in the fundamental box will have $\sim 1$-$10\%$ discrepancy between $\kstd$ and $\keff$. We will typically get a numerical GW ``noise'' with amplitude $A_{t,\rm noise}\sim 0.01$-$0.1 A_s$, where $A_s$ is the amplitude of the scalar metric perturbations. Thus, the relative error in the GW amplitude $A_t$ is $A_{t,\rm noise}/A_{t} \sim 0.01/r$-$0.1/r$, where $r\equiv A_t/A_s$ is the tensor-to-scalar ratio of metric perturbations. For a model with $r\lesssim 0.1$, we, in principle, cannot ignore this potential numerical noise due to imperfect TT projection, unless we understand that it will vanish in some way. In \refsec{subsec:chaotic} this scalar-tensor mixing problem due to imperfect TT projection will be further discussed with a concrete example.

\subsection{The sixth-order symplectic-Runge-Kutta hybrid integrator \label{subsec:symplectic}}

HLATTICE does not discretize the action along the temporal direction. Instead it uses an accurate sixth-order symplectic integrator to integrate the EOMs.

A symplectic integrator is designed to integrate a classical system with a Hamiltonian that has the form $\mathcal{H}(\vect{p},\vect{q})= \mathcal{K}(\vect{p}) + \mathcal{P}(\vect{q})$, where $\vect{q}$, $\vect{p}$, $\mathcal{P}(\vect{q})$ and  $\mathcal{K}(\vect{p})$ are, respectively, the general coordinates, the conjugate momenta, the potential energy and the kinetic energy. I have used the curled $\mathcal{H}$ to distinguish the Hamiltonian from the Hubble parameter $H$.

An arbitrary function $f(\vect{p},\vect{q})$ can be evolved with
\begin{equation}
\frac{df}{dt} = \opr{H} f\ , \label{poisson}
\end{equation}
where the functional operator $\opr{H}$ is defined as
\begin{equation}
\opr{H} f \equiv \{ f, \mathcal{H} \} \ ,
\end{equation}
with $\{\cdot, \cdot\}$ being the Poisson bracket. Similarly, we can define $\opr{K} \equiv \{ \cdot, \mathcal{K}\}$ and $\opr{P} \equiv \{ \cdot, \mathcal{P}\}$. Note that $\opr{H}=\opr{K}+\opr{P}$, and that $\opr{K}$ and $\opr{P}$ do not commute.

The solution of \refeq{poisson} can be formally written as
\begin{equation}
f(t+dt) = e^{\opr{H}dt}f(t)\ .
\end{equation}

The $n$th-order symplectic integrator is constructed by factorizing $e^{\opr{H}dt}= e^{(\opr{K}+\opr{P})dt}$ as
\begin{equation}
e^{\opr{H}dt} = e^{c_1\opr{K}dt}e^{d_1 \opr{P}dt}e^{c_2\opr{K}dt}e^{d_2 \opr{P}dt}... + O(dt^{n+1}) \ , \label{sympgeneral}
\end{equation}
where $c_1, d_1, c_2, d_2, ...$ are constant c-numbers.

The operators $\opr{K}$ and $\opr{P}$ can be regarded as, respectively, a Hamiltonian for free particles and that for ``inertialess'' particles with a potential. Consequently, the {\it exact} evolution of the system under $e^{\opr{K}dt}$ and $e^{\opr{P}dt}$ can be achieved numerically. More explicitly, the solutions are
\begin{equation}
e^{\opr{K}dt}
\begin{pmatrix}\vect{p} \\ \vect{q}
\end{pmatrix}
=
\begin{pmatrix}
\vect{p} \\
\vect{q} + \frac{\partial \mathcal{K}}{\partial \vect{p}} dt
\end{pmatrix}
\ , \label{Kpropg}
\end{equation}
and 
\begin{equation}
e^{\opr{P}dt}
\begin{pmatrix}\vect{p} \\ \vect{q}
\end{pmatrix}
=
\begin{pmatrix}
\vect{p} - \frac{\partial \mathcal{P}}{\partial \vect{q}} dt \\
\vect{q}
\end{pmatrix}
\ , \label{Ppropg}
\end{equation}
where $\frac{\partial \mathcal{K}}{\partial \vect{p}}$ should be understood as the vector $\left(\frac{\partial \mathcal{K}}{\partial p_1},\frac{\partial \mathcal{K}}{\partial p_2}, ...\right)$ and $\frac{\partial\mathcal{P}}{\partial \vect{q}}$ is defined in the same way.

Equations~(\ref{Kpropg}-\ref{Ppropg}) are {\it exact} for a finite $dt$. Using the right-hand side of Eq.~\refeq{sympgeneral} to evolve the system, we will only have an error term that scales as $dt^{n+1}$. (Strictly speaking, there are also machine round-off errors, which are $\sim 10^{-17}$ for Fortran double precision numbers that are used in HLATTICE.)

Because symplectic integrators are very stable, they are often used to study long-term evolution of many-body systems in astronomy and particle physics \cite{Saha1992,Forest2006}. The most well-known and oft-used symplectic integrator is the second-order one,
\begin{equation}
e^{\opr{H}dt} = e^{\opr{K} dt/2}e^{\opr{P}dt}e^{\opr{K}dt/2} + O(dt^3) \ ,
\end{equation}
 which is equivalent to the leapfrog algorithm used in other lattice codes \cite{Felder2008,Frolov2008,Sainio2010}. 

In HLATTICE, a modified sixth-order symplectic integrator is used. Before introducing the integrator, I will write down the discretized Hamiltonian of the scalar fields and gravity on the lattice.

Writing the integrals~(\ref{kfdef}-\ref{ggdef}) as the sums of the integrand on the lattice, we obtain the action of the discrete system described by $(m+6)n^3$ general coordinates, $\phi_\ell(i_1,i_2,i_3)$ and $\beta_{ij}(i_1,i_2,i_3)$ ($1\le \ell\le m$, $-n/2< i_1, i_2, i_3 \le n/2$), and by their time derivatives. Since rescaling the action by a constant factor does not change the EOMs of the system, we will drop the factor $\Delta^3$ in the discretized action.  

The conjugate momentum of $\phi_\ell(i_1,i_2,i_3)$ is
\begin{equation}
  \left.\Pi_{\phi_\ell}\right\vert_{(i_1,i_2,i_3)} = \left.e^{\beta/2}\dot\phi_\ell\right\vert_{(i_1,i_2,i_3)}\ ,
\end{equation}
and that of $\beta_{ij}(i_1,i_2,i_3)$ is
\begin{eqnarray}
\left.\Pi_{\beta_{ij}}\right\vert_{(i_1,i_2,i_3)} &=& \left.\frac{M_p^2}{4}e^{\beta/2}\left(2-\delta_{ij}\right)\left(\bdij{ij}-\dot\beta\delta_{ij}\right) \right\vert_{(i_1,i_2,i_3)}\ . \nonumber \\
\end{eqnarray}

Now we are ready to write down the Hamiltonian of the system, given by
\begin{equation}
\mathcal{H} =\mathcal{K}_1 + \mathcal{K}_2  + \mathcal{P}\ , \label{totH}
\end{equation}
where $\mathcal{K}_1$ is the kinetic energy of the scalar fields and the sum of off-diagonal terms in the ``kinetic energy'' of gravity, 
\begin{equation}
\mathcal{K}_1 = \sum_{\rm lattice} e^{-\beta/2}\left[\frac{\Pi^2_{\phi_\ell}}{2} + \frac{1}{M_p^2} \left(\Pi_{\bij{23}}^2+\Pi_{\bij{31}}^2+\Pi_{\bij{12}}^2\right)\right]\ ;
\end{equation}
$\mathcal{K}_2$ is the sum of diagonal terms in the ``kinetic energy'' of gravity
\begin{equation}
\mathcal{K}_2 = \frac{1}{M_p^2} \sum_{\rm lattice} e^{-\beta/2}\left[2\sum_{i=1}^3\Pi^2_{\bij{ii}}-\left(\sum_{i=1}^3\Pi_{\bij{ii}}\right)^2\right] \ ;
\end{equation}
and $\mathcal{P}$ is the sum of all gradient and potential energy terms, 
\begin{eqnarray}
\mathcal{P} &=& \sum_{\rm lattice} e^{\beta/2}\Bigl[ V\left(\phi_1,\phi_2,...,\phi_m\right) + \frac{1}{2}g^{ij}\partial_i\phi_\ell\partial_j\phi_\ell\Bigr] \nonumber \\
&& + \frac{M_p^2}{4n}\left(\sum_{\rm lattice} e^{\beta/2}\right)^{1/3}\nonumber \\
&& \times \Bigl[\sum_{\rm lattice} \left( \bij{23,1}^2 + \bij{31,2}^2 + \bij{12,3}^2 \right. \nonumber  \\ 
&& - 2 \bij{23,1} \bij{31,2}  - 2 \bij{31,2} \bij{12,3} - 2\bij{12,3}\bij{23,1}\nonumber \\
&& - \bij{22,1} \bij{33,1} - \bij{33,2} \bij{11,2}-  \bij{11,3}\bij{22,3} \nonumber \\
&& \left.  + 2\bij{23,2} \bij{11,3} + 2 \bij{31,3}\bij{22,1} + 2 \bij{12,1}\bij{33,2}\right) \Bigr]\ ,
\end{eqnarray}
with $g^{ij}$ given by Eqs.~\refeq{gijeq}.

The symplectic integrators found in earlier works \cite{Ruth1983,Kinoshita1990,Yoshida1990} can not be directly used here. This is due to two problems: (i) $\opr{H}\equiv \{\cdot, \mathcal{H} \}$ contains three noncommutative operators $\opr{K}_1 \equiv \{\cdot, \mathcal{K}_1\}$, $\opr{K}_2  \equiv \{\cdot, \mathcal{K}_2\}$ and $\opr{P} \equiv \{\cdot, \mathcal{P}\}$, while in the literature only two-term symplectic factorization formulas are given; (ii) $\opr{K}_2$ is noncanonical, as it depends on both $\bij{11}$ and $\Pi_{\bij{11}}$.

The first problem in principle could be solved by iterative factorization. We can treat $\opr{K}_1+\opr{K}_2$ as one operator, factorize $e^{(\opr{K}_1+\opr{K}_2)dt + \opr{P}dt}$, and finally factorize each factor that contains $\opr{K}_1+\opr{K}_2$. This procedure, however, is not optimal, and leads to a factorization formula with hundreds of factors (for sixth or higher order). Indeed a much simpler symplectic factorization exists, as I give below.

For arbitrary operators $\opr{A},\ \opr{B},\ \opr{C}$, commuting or not, $e^{(\opr{A}+\opr{B}+\opr{C})dt}$ can be factorized as
\begin{eqnarray}
e^{(\opr{A}+\opr{B}+\opr{C})dt} &=& e^{c_3\opr{A}dt/2}e^{c_3\opr{B}dt/2}e^{c_3\opr{C}dt}e^{c_3\opr{B}dt/2}e^{(c_3+c_2)\opr{A}dt/2}\nonumber \\
&& \times \, e^{c_2\opr{B}dt/2}e^{c_2\opr{C}dt}e^{c_2\opr{B}dt/2}e^{(c_2+c_1)\opr{A}dt/2} \nonumber \\
&& \times \, e^{c_1\opr{B}dt/2}e^{c_1\opr{C}dt}e^{c_1\opr{B}dt/2}e^{(c_1+c_0)\opr{A}dt/2} \nonumber \\
&& \times \, e^{c_0\opr{B}dt/2}e^{c_0\opr{C}dt}e^{c_0\opr{B}dt/2}e^{(c_0+c_1)\opr{A}dt/2} \nonumber \\
&& \times \, e^{c_1\opr{B}dt/2}e^{c_1\opr{C}dt}e^{c_1\opr{B}dt/2}e^{(c_1+c_2)\opr{A}dt/2} \nonumber \\
&& \times \, e^{c_2\opr{B}dt/2}e^{c_2\opr{C}dt}e^{c_2\opr{B}dt/2}e^{(c_2+c_3)\opr{A}dt/2} \nonumber \\
&& \times \, e^{c_3\opr{B}dt/2}e^{c_3\opr{C}dt}e^{c_3\opr{B}dt/2}e^{c_3\opr{A}dt/2} \nonumber \\
&& + O(dt^7)\ , \label{symp6}
\end{eqnarray}
where
\begin{eqnarray}
c_1 &=& -1.17767998417887 \ ,\nonumber \\
c_2 &=& 0.235573213359357 \ ,\nonumber \\
c_3 &=& 0.784513610477560 \ ,\nonumber \\
c_0 &=& 1-2(c_1+c_2+c_3) \ . \label{symp6coef}
\end{eqnarray}
Eq.~\refeq{symp6} can be checked by expanding both sides up to sixth-order in $dt$. A Python script doing this tedious but straightforward job can be downloaded from {\small http://www.cita.utoronto.ca/$\sim$zqhuang/work/symp6.py\,}.

The symplectic factorization is not unique. For example, one can interchange $\opr{A}$ and $\opr{B}$ in Eq.~\refeq{symp6} to generate another correct sixth-order symplectic factorization with a different $O(dt^7)$ residual term. 

While the proof of a symplectic factorization is always trivial, the technique to search for one is not. A general approach is to assume a form of the factorization with a few undetermined coefficients and solve for these coefficients by requiring the exact cancellation of lower-order terms. For a high-order symplectic factorization, these coefficients often need to be solved numerically, by using the Monte-Carlo technique to search for a solution in the high-dimensional parameter space. This is how I obtained Eq.~\refeq{symp6}. If $\opr{C}=0$,  Eq.~\refeq{symp6} degrades to a two-term symplectic factorization formula containing the same set of coefficients in Eq.~\refeq{symp6coef}, which, indeed, was found two decades ago in Ref.~\cite{Yoshida1990}. 

Because factors containing $\opr{B}$ appear more frequently in Eq.~\refeq{symp6}, I let $\opr{B}=\opr{K}_1$, whose numerical evaluation is less expensive. Letting $\opr{A}=\opr{K}_2$ and $\opr{C}=\opr{P}$ and noticing that the c-numbers can always be absorbed into $dt$, we now only need to write down the explicit algorithm to evolve the general coordinates $\phi_\ell$ and $\beta_{ij}$ and their conjugate momenta under the operators $e^{\opr{K}_1dt}$, $e^{\opr{K}_2dt}$, and $e^{\opr{P}dt}$. For canonical operators $e^{\opr{K}_1dt}$ this is straightforward. We take $\mathcal{K}_1$ as the Hamiltonian and write down the Hamiltonian equations:
\begin{equation}
e^{\opr{K_1}dt}
\begin{pmatrix}
\left.\phi_\ell\right\vert_{i_1,i_2,i_3} \\
\left.\Pi_{\phi_\ell}\right\vert_{i_1,i_2,i_3} \\
\left.\beta_{11}\right\vert_{i_1,i_2,i_3} \\
\left.\beta_{22}\right\vert_{i_1,i_2,i_3} \\
\left.\beta_{33}\right\vert_{i_1,i_2,i_3} \\
\left.\beta_{23}\right\vert_{i_1,i_2,i_3} \\
\left.\beta_{31}\right\vert_{i_1,i_2,i_3} \\
\left.\beta_{12}\right\vert_{i_1,i_2,i_3} \\
\left.\Pi_{\beta_{11}}\right\vert_{i_1,i_2,i_3} \\
\left.\Pi_{\beta_{22}}\right\vert_{i_1,i_2,i_3} \\
\left.\Pi_{\beta_{33}}\right\vert_{i_1,i_2,i_3} \\
\left.\Pi_{\beta_{23}}\right\vert_{i_1,i_2,i_3} \\
\left.\Pi_{\beta_{31}}\right\vert_{i_1,i_2,i_3} \\
\left.\Pi_{\beta_{12}}\right\vert_{i_1,i_2,i_3} \\
\end{pmatrix}
=
\begin{pmatrix}
\left.\left(\phi_\ell + e^{-\beta/2}\Pi_{\phi_\ell}dt\right) \right\vert_{i_1,i_2,i_3}\\
\left.\Pi_{\phi_\ell}\right\vert_{i_1,i_2,i_3} \\
\left.\beta_{11}\right\vert_{i_1,i_2,i_3} \\
\left.\beta_{22}\right\vert_{i_1,i_2,i_3} \\
\left.\beta_{33}\right\vert_{i_1,i_2,i_3} \\
\left.\left(\beta_{23}+\frac{2e^{-\beta/2}}{M_p^2}\Pi_{\beta_{23}}dt\right) \right\vert_{i_1,i_2,i_3}\\
\left.\left(\beta_{31}+\frac{2e^{-\beta/2}}{M_p^2}\Pi_{\beta_{31}}dt\right) \right\vert_{i_1,i_2,i_3}\\
\left.\left(\beta_{12}+\frac{2e^{-\beta/2}}{M_p^2}\Pi_{\beta_{12}}dt\right) \right\vert_{i_1,i_2,i_3}\\
\left.\Pi_{\beta_{11}}+\frac{\mathcal{K}_1}{2}dt\right\vert_{i_1,i_2,i_3}   \\
\left.\Pi_{\beta_{22}}+\frac{\mathcal{K}_1}{2}dt\right\vert_{i_1,i_2,i_3} \\
\left.\Pi_{\beta_{33}}+\frac{\mathcal{K}_1}{2}dt\right\vert_{i_1,i_2,i_3} \\
\left.\Pi_{\beta_{23}}\right\vert_{i_1,i_2,i_3} \\
\left.\Pi_{\beta_{31}}\right\vert_{i_1,i_2,i_3} \\
\left.\Pi_{\beta_{12}}\right\vert_{i_1,i_2,i_3} \\
\end{pmatrix}
\ , \label{HLK1propg}
\end{equation}
Since the quantities used to evolve the system, $\beta$, $\Pi_{\phi_\ell}$, $\Pi_{\beta_{23}}$, $\Pi_{\beta_{31}}$, $\Pi_{\beta_{12}}$ and $\mathcal{K}_1$ all remain unchanged under $e^{\opr{K}_1dt}$, the above algorithm can be applied for finite $dt$ without any ambiguity. For example, one does not need to ask whether $\mathcal{K}_1$ on the right-hand side of Eq.~\refeq{HLK1propg} is evaluated at time $t$ or $t+dt$. 

The same procedure can be applied to derive the lattice-version EOMs under the canonical operator $e^{\opr{P}dt}$, by taking $\mathcal{P}$ as the Hamiltonian and writing down the Hamiltonian equations. Because of the complexity of $G_f$ and $G_g$, the final result is not human-readable. The interested reader is referred to the macro files in the released HLATTICE package, where the EOMs are defined via a series of compiler-preprocessor macros.

Under the operator $e^{\opr{K}_2dt}$, the EOMs are
\begin{equation}
e^{\opr{K}_2dt}
\begin{pmatrix}
\beta_{11} \\
\beta_{22} \\
\beta_{33}\\
\Pi_{\beta_{11}}\\ 
\Pi_{\beta_{22}}\\
\Pi_{\beta_{33}} 
\end{pmatrix}
=
\begin{pmatrix}
\beta_{11}+\frac{2e^{-\beta/2}}{M_p^2}\left(\Pi_{\beta_{11}}-\Pi_{\beta_{22}}-\Pi_{\beta_{33}}\right)dt \\
\beta_{22}+\frac{2e^{-\beta/2}}{M_p^2}\left(\Pi_{\beta_{22}}-\Pi_{\beta_{33}}-\Pi_{\beta_{11}}\right)dt \\
\beta_{33}+\frac{2e^{-\beta/2}}{M_p^2}\left(\Pi_{\beta_{33}}-\Pi_{\beta_{11}}-\Pi_{\beta_{22}}\right)dt \\
\Pi_{\beta_{11}}+\frac{\mathcal{K}_2}{2} dt\\
\Pi_{\beta_{22}}+\frac{\mathcal{K}_2}{2} dt\\
\Pi_{\beta_{33}}+\frac{\mathcal{K}_2}{2} dt
\end{pmatrix}\ , \label{HLK2propg}
\end{equation}
where the configuration-space label $\vert_{i_1,i_2,i_3}$ is omitted on both sides. The rest of the general coordinates and conjugate momenta do not change under $e^{\opr{K}_2dt}$.

However, due to the noncanonicality of $\opr{K}_2$, the exact solution for Eq.~\refeq{HLK2propg} with finite $dt$ cannot be achieved. This is because $\mathcal{K}_2$ depends on both $\beta_{11}$,  $\beta_{22}$,  $\beta_{33}$ and their conjugate momenta. Consequently $\beta$, $\Pi_{\beta_{11}}$, $\Pi_{\beta_{22}}$, $\Pi_{\beta_{33}}$ and $\mathcal{K}_2$ are all dynamical on the right-hand side of Eq.~\refeq{HLK2propg}. Thus an algebraic solution does not exist for a finite $dt$. 
To solve Eq.~\refeq{HLK2propg} I use a fourth-order Runge-Kutta integrator {\it with a smaller time step $dt'\ll dt$}. This Runge-Kutta subintegrator, unlike a global one, does not cost extra memory, because the operator $\opr{K}_2$ is local (does not contain interactions between different grid points). Here we are solving $n^3$ {\it independent} sets of ODE, with each set containing six coupled ODE. If a global Runge-Kutta integrator had been used, the task would then be solving $6n^3$ {\it all-coupled} ODEs, which is numerically much more expensive. Since this step is numerically cheap, I can make $dt'$ sufficiently small so that the $O(dt'^5)$ error from the fourth-order Runge-Kutta integrator does not spoil the global $O(dt^6)$ accuracy. In HLATTICE $dt/dt'$ is an adjustable parameter, whose default value is set to be $10$.

\begin{figure}
\includegraphics[width=\figurewidth]{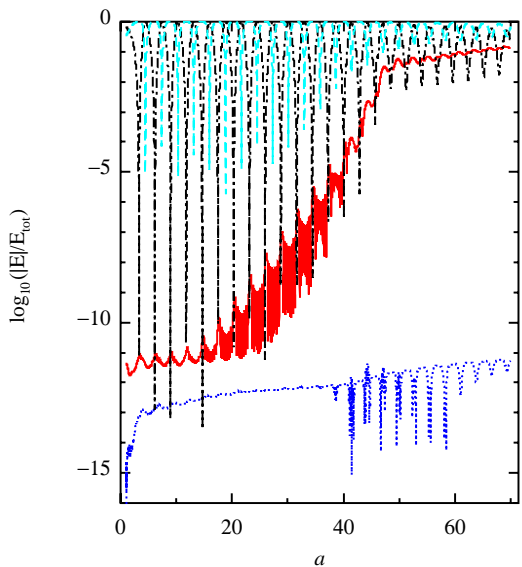}
\caption{\label{figEfluc} Lattice simulation for the preheating model, $V=\frac{\lambda}{4}\phi^4 + \frac{1}{2}g^2\phi^2\chi^2$, using HLATTICE, where $\lambda=10^{-13}$ and $g^2/\lambda = 200$. The simulation is initiated at the end of inflation, where I define $a=1$ and choose the box size $L=20H^{-1}$. The solid red line is $\log_{10}(E_{\rm grad}/E_{\rm tot})$, where $E_{\rm grad}$ is the mean gradient energy, and $E_{\rm tot}$ is the mean total energy. The dot-dashed black line and dashed cyan line are $\log_{10} (E_{\rm pot}/E_{\rm tot})$ and $\log_{10} (E_{\rm kin}/E_{\rm tot})$, where $E_{\rm pot}$ and $E_{\rm kin}$ are the mean potential energy and the mean kinetic energy, respectively.  The dotted blue line, $\log_{10}|3H^2M_p^2/E_{\rm tot}-1|$, shows that the constraint equation is satisfied at the $10^{-12}$ level.}
\end{figure}

Both the $O(dt'^5)$ error from the Runge-Kutta subintegrator and the $O(dt^7)$ term in Eq.~\refeq{symp6} can be made very small by shrinking $dt$ by a factor of a few. This symplectic-Runge-Kutta hybrid integrator can thus be made very accurate without much additional computational cost. In \reffig{figEfluc} I show a simulation done on an eight-core desktop PC in about half an hour. (All eight cores are used, as HLATTICE is an OpenMP parallelized code.) 
The fractional energy noises are suppressed to $\lesssim 10^{-12}$. Such noises are $\sim 10^{-5}$-$10^{-3}$ in other public available lattice codes and cannot be suppressed much by shrinking $dt$, because the error term behaves as $O(dt^3)$ in those codes. Here, for illustration purposes, the metric perturbations are turned off, and a low resolution $n=64$ is chosen. It is about 10 times more expensive to include metric perturbations. Hence a simulation with metric perturbations typically takes about $5(\frac{n}{64})^3(\frac{8}{\rm \# \ of\ cores})\,{\rm h}$ on a desktop PC, assuming the simulation can be done with roughly the same number of ($\sim 50,000$) time steps. However, practically a simulation including metric perturbations is often memory-limited. A simulation with $n\ge 256$, including metric perturbations, either crashes (on low-memory machines) or becomes very slow on most computer architectures. This deficiency could be partially cured by using distributed memory, at the price of frequent boundary data exchange. I leave the development of a MPI-parallelized (distributed-memory) version of HLATTICE to future work.

In HLATTICE a few lower-order symplectic integrators are also given as alternative options. The computational cost can be reduced if a lower accuracy is tolerable.

\subsection{The gauge choice problem}\label{subsec:gauge}

I have written HLATTICE in synchronous gauge for practical convenience. In synchronous gauge the gauge condition $g_{00}=g_{0i}=0$ is local. In other gauges the ten metric variables $g_{\mu\nu}$ are constrained by four global constraint equations, which have to be solved at every time step in order to eliminate the four gauge degrees of freedom. This is computationally expensive. Another option is to keep all ten variables $g_{\mu\nu}$ on the lattice and evolve them with Hamiltonian equations and four external constraint equations defined by the gauge condition. However, a symplectic integrator cannot deal with external constraint equations, at least not in a trivial way. Thus, writing HLATTICE in other gauges is a difficult problem. Nevertheless, I will discuss the theoretical aspect of a generic gauge choices without implementing it in a numerical code.

At first glance, gauge invariance must be broken when we use the exact energy-momentum tensor $T_{\mu\nu}$ on the right-hand side of the Einstein equation, while keeping only the first-order terms in the Einstein tensor $G_{\mu\nu}$ on the left-hand side:
\begin{equation}
G_{\mu\nu}^{(0)} + G_{\mu\nu}^{(1)} = \frac{1}{M_p^2} T_{\mu\nu}^{\rm exact} \ . \label{1stEin}
\end{equation}
Here $G_{\mu\nu}^{(0)}$ is the background quantity, and $G^{(1)}_{\mu\nu}$ contains linear terms of metric perturbations $\delta g_{\mu\nu}$ (or spacetime derivatives of them). Analogously, we can define $G^{(2)}_{\mu\nu}$, $G_{\mu\nu}^{(3)}$, etc. In the presence of small-scale nonlinearity in $T_{\mu\nu}^{\rm exact}$, the spacetime derivatives of the metric perturbations are not necessarily small, but we still assume $\delta g_{\mu\nu}$ to be small. (Collapsed objects like blackholes are not considered here.) For example, in the late-time Universe around a dark matter halo, the Laplacian of the Newtonian potential $\nabla^2\Phi_N$ can be larger than the ``zero-th order'' quantity $H^2$, while $\Phi_N$ remains $\sim 10^{-5}$. Because $G_{\mu\nu}^{(1)}$ contains $\partial^2 g_{\mu\nu}$ terms, it can be comparable to $G_{\mu\nu}^{(0)}$. The second-order $G_{\mu\nu}^{(2)}$, which for dimensional reason does not contain more derivatives, is suppressed by one more power of $\delta g_{\mu\nu}$. Hence it can be ignored.

Now, let us perform an infinitesimal coordinate transformation $x^{\mu} \rightarrow x^{\mu} - \epsilon \xi^{\mu}$. The two sides of Eq.~\refeq{1stEin} transform as
\begin{eqnarray}
G^{(0)}_{\mu\nu} + G^{(1)}_{\mu\nu} &\rightarrow& G^{(0)}_{\mu\nu} + G^{(1)}_{\mu\nu} + \epsilon L_{\xi} G^{(0)}_{\mu\nu} +  \epsilon L_{\xi} G^{(1)}_{\mu\nu} \label{Gtrans}\ , \\
T_{\mu\nu}^{\rm exact} &\rightarrow& T_{\mu\nu}^{\rm exact} + \epsilon L_{\xi} T_{\mu\nu}^{\rm exact}\label{Ttrans} \ ,
\end{eqnarray}
where $L_{\xi}$ is the Lie derivative along $\xi$.

If we had followed the (usual) first-order gauge transformation formulas for the metric \cite{Malik2009}, we would have discarded the last ``second-order'' term on the right-hand side of \refeq{Gtrans}. However, we know that we should not do so, as  on small scales  $ \epsilon L_{\xi} G^{(1)}_{\mu\nu}$ is of the same order of $\epsilon L_{\xi} G^{(0)}_{\mu\nu}$. Both terms are needed to match the $ \epsilon L_{\xi} T_{\mu\nu}^{\rm exact}$ term in Eq.~\refeq{Ttrans}, otherwise the Einstein equation in the new gauge would not be equivalent to the one in the old gauge.

In summary, to be self-consistent we need to use new gauge transformation rules where spacetime derivatives of $\delta g_{\mu\nu}$ are treated as zeroth-order quantities. This then leads to a problem: $\delta g_{\mu\nu}$ themselves may no longer be small after a gauge transformation. Physically, this means that there are ``optimal gauges'' where metric perturbations remain small. This is not surprising, as we expect, for example, that uniform energy gauge would fail in the presence of inhomogeneous matter.

The question for HLATTICE is then whether the synchronous gauge is one of the ``optimal gauges''. The answer depends on how long we want to evolve the system and how inhomogeneous the scalar fields are. Empirically when $h_{ij}$ approaches $O(1)$ we should stop the simulation, because in this case we are using rather bad approximations. This, however, does not necessarily indicate a strong gravitational field. In synchronous gauge one has the freedom to choose arbitrary spatial coordinates. This corresponds to a nonphysical gauge mode, whose amplitude (but not its time derivative) can be eliminated {\it at any given moment} by a proper gauge transformation $x^i\rightarrow x^i+\zeta^i(\vect{x})$. HLATTICE~V2.0 includes an option of using adaptive spatial coordinates (i.e. eliminating the gauge mode in every few steps). Currently this option is labeled as a ``beta version'' (a trial version). Its numerical stability will be studied in my future work. 

A conjecture is that Newtonian gauge is more optimal, since the Newtonian potentials are all physical. I also leave the rewriting of HLATTICE in Newtonian gauge for future work, if a proper integrator can be found for this gauge.

I end this section by summarizing the current HLATTICE configuration options in Table~\ref{tblHLoptions}.
\begin{table*}
\caption{Available Options in HLATTICE\label{tblHLoptions}. S2-6 stands for the option of using the second, fourth and sixth symplectic integrators; GW stands for the option to calculate gravity waves; $\tau$ stands for the option of using conformal time (the default is physical time); $\vkeff$ ($\vkstd$) stands for the option of using $\vkeff$ ($\vkstd$) to construct the TT projector.  }
\begin{tabular}{p{0.32\textwidth}|l|l|l}
\hline
\hspace{0.12\textwidth}   discretization  scheme &\  Eq.~\refeq{LEnabla2}\ \ \ & Eq.~\refeq{derv1}\ \ \ & Eq.~\refeq{derv2} \\

metric  & & & \\
\hline
Minkowski, no perturbations  &\ S2-6; GW; $\vkstd$ & S2-6; GW; $\vkeff$; $\vkstd$  & S2-6; GW; $\vkeff$; $\vkstd$ \\
\hline
FRW, no perturbations  &\  S2-6; GW; $\tau$; $\vkstd$  &  S2-6; GW; $\tau$; $\vkeff$; $\vkstd$ &  S2-6; GW; $\tau$; $\vkeff$; $\vkstd$ \\
\hline
FRW, with perturbations, synchronous gauge, fixed spatial coordinates &\ DISABLED & S2-6; GW; $\vkeff$; $\vkstd$ &  S2-6; GW; $\vkeff$; $\vkstd$ \\
\hline
FRW, with perturbations, synchronous gauge, adaptive spatial coordinates &\ DISABLED &  S2-6; GW; $\vkeff$ &  S2-6; GW; $\vkeff$\\
\hline
\end{tabular}
\end{table*}

\section{Gravity waves from preheating \label{sec:GW}}

\subsection{Gravity waves from tachyonic preheating after hybrid inflation \label{subsec:hybrid}}
In this section I use HLATTICE to calculate gravitational waves from tachyonic preheating after hybrid inflation \cite{Garcia2008,Dufaux2009}. Following \cite{Garcia2008,Dufaux2009}, I assume a real inflaton field $\phi$ and a complex field $\sigma = \sigma_1+ \ii \sigma_2$. The potential reads
\begin{equation}
V = \frac{1}{4}\lambda (\sigma^2 - \upsilon^2)^2 + \frac{1}{2}g^2 \phi^2 \sigma^2\ ,
\end{equation}
where $\sigma^2 \equiv \sigma_1^2 + \sigma_2^2$.

For illustration purpose, I fix the parameters that have been used in Figure~4 of \cite{Dufaux2009}: $\lambda=10^{-14}, g^2/\lambda=2, \upsilon=10^{-3}\sqrt{8\pi}M_p$, and $\frac{g\dot\phi}{\lambda\upsilon^2}\vert_{\phi=\phi_c}=10^{-5}$, where $\phi_c\equiv \sqrt{\lambda}\upsilon/g$ is the critical point where inflation ends. At the beginning of the simulation, the initial metric perturbations are all set to zero and $a$ is defined to be $1$. The $\sigma$ field is initialized with random Gaussian fluctuations with ``vacuum-fluctuation'' amplitude $|\sigma_{1,k}|^2=|\sigma_{2,k}|^2=1/(2\omega_k)$ and $|\dot\sigma_{1,k}|^2=|\dot\sigma_{2,k}|^2=\omega_k/2$, where $\omega_k \equiv \sqrt{k^2+m_{\sigma}^2} =k$, as $\sigma$ is massless at the beginning of simulation. This is an approximation using the classical lattice simulation to mimic how these quantum vacuum modes become classical due to the tachyonic instability. Since the growth of $\delta\sigma$ is exponential, this approximation is expected to be good. Another problem is that, physically, gravity should not respond to these unrenormalized vacuum modes, while on the classical lattice it does respond to them. However, since $\sigma$ is light and the lattice UV cutoff is not too high, the nonphysical $h_{ij}$ excited by these initial $\sigma$ fluctuations is negligible. The inflaton $\phi$ is set to be initially homogeneous, for two reasons: (i) At the beginning, the dominating physics is the tachyonic growth of $\sigma$ fluctuations. Later $\phi$ fluctuations are excited and enhanced by the inhomogenous $\chi$ field. The vacuum flutuations in $\phi$ remain irrelevant here. (ii) the renormalization problem is more severe for the $\phi$ field, as at the beginning the energy fluctuations are more sensitive to $\phi$ fluctuations.

\begin{figure}
\includegraphics[width=\figurewidth]{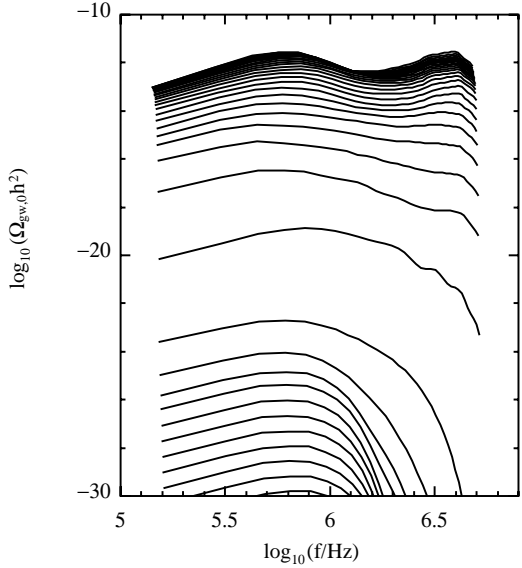}
\caption{\label{figgw} Gravity waves from tachyonic preheating after hybrid inflation calculated using HLATTICE V1.0 with metric perturbations and fixed spatial coordinates. The simulation resolution is $n=128$. The model and parameters are the same as those that have been used in Figure~4 of \cite{Dufaux2009}. Here $\Omega_{\rm gw, 0}$ is the fractional energy of GW (per efold in frequency $f$) that would be observed today (assuming radiation domination after preheating); $h$ is the current Hubble parameter in unit of $100\, {\rm km\,s^{-1}\,Mpc^{-1}}$. The outputs are plotted from bottom to top per unit time step $d t = 0.00313 H_{\phi=\phi_c}^{-1}$, where  $\phi_c$ is the critical point where inflation ends. The lattice simulation is initiated at $\phi = 0.9975\phi_c$, where I define $a\equiv 1$ and the box size of simulation $L=0.8H_{\phi=\phi_c}^{-1}$.}
\end{figure}

The GW energy spectrum computed using HLATTICE is shown in \reffig{figgw}. The fractional energy of GWs per e-fold, $\Omega_{\rm gw}$, is defined as 
\begin{equation}
\Omega_{\rm gw} \equiv \frac{1}{\rho_{\rm crit}}\frac{d \rho_{\rm gw}}{ d\ln f}\ , \label{omgwdef}
\end{equation}
where $f$ is the GW frequency and $\rho_{\rm gw}$ is the energy density of the GW. Here the critical density, defined as $\rho_{\rm crit}\equiv 3H^2M_p^2$, in a spatially flat Universe is the same as the mean energy density. I have converted the GW energy spectrum to present-day observables in \reffig{figgw}. $\Omega_{\rm gw,0}$ is defined by replacing all quantities by today's observables in Eq.~\refeq{omgwdef}. I have used Eqs.~(35-36) in Ref.~\cite{Dufaux2009} to convert the simulation output to present-day observables. See also Ref.~\cite{Garcia2008} for a more detailed derivation of these formulas.

In this model, GW are produced during two stages.

In the first stage, $\phi$ can be approximated as $\phi = \phi_c - \dot \phi_c t$. The mass square of the $\sigma$ field is approximately 
\begin{equation}
m_{\sigma}^2 \approx - 2g^2\phi_c\dot\phi_c t \ . \label{m2sigma}
\end{equation}
The infrared modes $k < g\sqrt{2\phi_c\dot\phi_c t}$ first start to grow. As $t$ increases, more and more modes become tachyonic and begin to grow. Bubbles of $\sigma$ field are created in this process, producing gravity waves on roughly the same scales \cite{Dufaux2009}. This can be seen from the lower part of \reffig{figgw}. 

In the second stage, $\phi$ condensate is broken due to the coupling between $\phi$ and $\sigma$. The estimation \refeq{m2sigma} is no longer valid. The typical scales of inhomogeneity rapidly shift towards $k\sim g\phi_c$, producing GW waves on these scales. At low $k$ the GW spectrum saturates at a stationary level, as shown in the upper part of \reffig{figgw}. The saturated low-frequency part of GW is what we are interested in, as it can potentially be observed with future GW probes such as BBO \cite{Garcia2008,Dufaux2009}.

\begin{figure}
\includegraphics[width=\figurewidth]{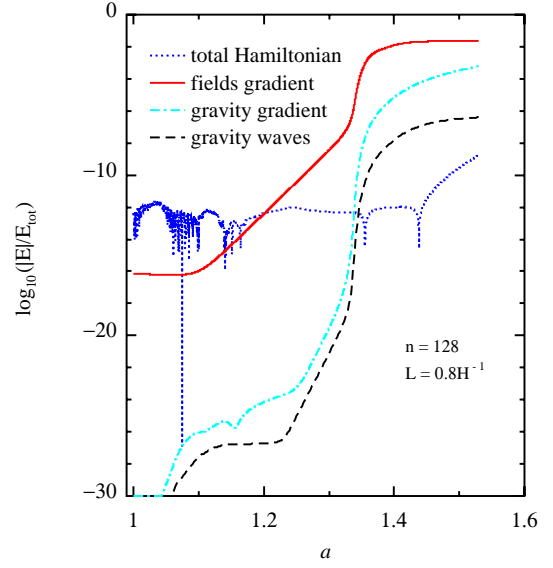}
\caption{\label{figgg} The comparison between energy carried by GW and numerical energy noise. The simulation is the same one as that described in \reffig{figgw}. All the quantities are evaluated at the end of the simulation where $a\approx 1.8$. The solid red line is $\log_{10}(E_{\rm grad, fields}/E_{\rm tot, fields})$, where $E_{\rm grad, fields}$ is the total gradient energy of $\phi$ and $\sigma$ fields, and $E_{\rm tot, fields}$ the total energy carried by them. The dot-dashed cyan line is $\log_{10}(|G_g|/E_{\rm tot, fields})$, where $G_g$ is the ``gradient energy'' of gravity defined in \refeq{ggdef}. The dashed black line is $\log_{10}(E_{\rm GW}/E_{\rm tot, fields})$, where $E_{\rm GW}$ is the energy carried by GW. The dotted blue line is $\log_{10}(\mathcal{H}/E_{\rm tot, fields})$, with $\mathcal{H}$ being the total Hamiltonian given by Eq.~\refeq{totH}.}
\end{figure}

When gravity is included, the constraint equation is always $\mathcal{H}=0$ regardless of the matter content of the system. This constraint equation can be used to estimate the numerical energy noises. In \reffig{figgg} I plot the energy carried by GW and the total Hamiltonian $\mathcal{H}$, both divided by the energy carried by the scalar fields for comparison. The contribution of energy carried by GW saturates to $\sim 10^{-6}$, while the final numerical energy noise is $\sim 10^{-9}$. In other words, energy conservation has been checked at the level of about 0.1\% of the energy carried by GW.

Comparing the ``gradient energy'' of gravity $G_g$ shown in \reffig{figgg} to the energy carried by GW, we can estimate $r(k) \sim 10^{-2}$ for the dominating modes. In this example, the relative contribution of GW due to an imperfect TT projector could be as high as $\sim 0.1/r$, i.e., about $10$ times larger than the physical GW spectrum shown in \reffig{figgw}. This provides a possible explanation for why the GW spectra found in previous works (see \cite{Dufaux2009, Bastero2010} and references therein) are generally larger than what I have obtained using HLATTICE. However, the discrepancy could also be due to the fact that I have included the expansion of the Universe, which is ignored in Ref.~\cite{Dufaux2009}. Finally, due to the memory limitation, I cannot achieve the same high spatial resolution as in Ref.~\cite{Dufaux2009}, which the authors found is also important for this model. 

The feedback from metric perturbations, however, is found to be irrelevant for this model. No significant difference has been found in the power spectra of the scalar fields between simulations with and without metric perturbations.

\subsection{Gravity Waves from preheating after chaotic inflation \label{subsec:chaotic}}

The example we have shown in the previous section, although being observationally interesting, is physically complicated. The GW energy spectrum has two peaks due to different physics at two stages. Mode-mode coupling in a wide range of scales becomes important in the nonlinear regime, which can hardly be captured by a simulation with $n=128$. The result shown in \reffig{figgw} needs to be further studied with a simulation with much higher resolution if HLATTICE can be MPI-parallelized.

\begin{figure}
\includegraphics[width=\figurewidth]{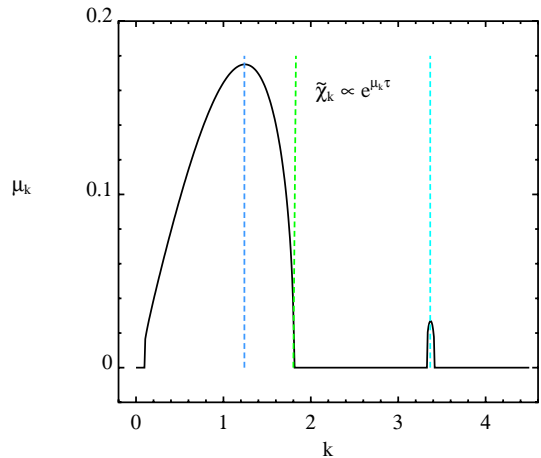}
\caption{The Floquet exponent $\mu_k(k)$ for $\tilde{\chi}_k (\tau)$ in the preheating model $V(\phi,\chi) = \frac{\lambda}{4}\phi^4+\frac{g^2}{2}\phi^2\chi^2$ with $g^2/\lambda=120$, where $\tau$ is the conformal time in units of $1/(\sqrt{\lambda}\tilde{\phi}_{\rm max})$ and $k$ is the comoving wave number in units of $\sqrt{\lambda}\tilde{\phi}_{\rm max}$. Here $\tilde{\chi}\equiv a\chi$ and $\tilde {\phi}\equiv a\phi$ are the conformal field values; $\tilde{\phi}_{\rm max}$ is the amplitude of background $\tilde{\phi}$ oscillations in the linear regime. The dashed sky-blue, green and cyan lines are three characteristic scales that will be  compared to the simulation results.\label{muk}}
\end{figure}

To better understand the scalar-tensor mixing problem, it is better to take a simple example that has been well studied and well understood, and can be fully captured by a simulation with resolution $n=128$. Here I take the example of preheating after chaotic inflation,
\begin{equation}
V(\phi,\chi) = \frac{\lambda}{4}\phi^4+\frac{g^2}{2}\phi^2\chi^2\ . \label{lf4}
\end{equation}
The parametric-resonance bands and Floquet exponents for this model have been studied in detail in Ref.~\cite{Greene1997}.  I choose the parameter $\lambda=10^{-14}$ and $g^2/\lambda = 120$. This set of parameters has been studied in Refs.~\cite{Garcia2008, Easther2007, Dufaux2007}. For $g^2/\lambda =120$ the Floquet exponent $\mu_k$ is shown in Figure~\ref{muk}. The dominating mode and the boundary of the first resonance band, and the dominating mode in the second resonance band are labeled with dashed sky-blue, green and cyan lines, respectively.

 I initialize the fields with ``vacuum-amplitudes'' random Gaussian fluctuations with a cutoff $|j_1|, |j_2|, |j_3| < 38$ (where $\vkeff$ does not significantly differ from $\vkstd$). This covers the resonance band shown in Figure~\ref{muk}. The simulation resolution is $n=128$. The box size at the beginning of the simulation is $L=n\Delta=20/(\sqrt{\lambda}\phi_{\rm ini}) = 13.87H^{-1}_{\rm ini}$, where $\phi_{\rm ini}=1.714M_p$ is the LATTICEEASY default initial background value of $\phi$, and $H_{\rm ini}$ is the initial Hubble parameter.  The scale factor $a$ is defined to be $1$ at the beginning of the simulation. Here, in order to compare my result with the previous works, I have used the same simulation configurations that have been used in Refs.~\cite{Garcia2008, Easther2007, Dufaux2007}. For this model, it is better to use conformal time as a time variable. In this simulation I thus do not include the backreaction of metric fluctuaions, which requires the coordinate $t$ to be the physical time.

\begin{figure}
\includegraphics[width=\figurewidth]{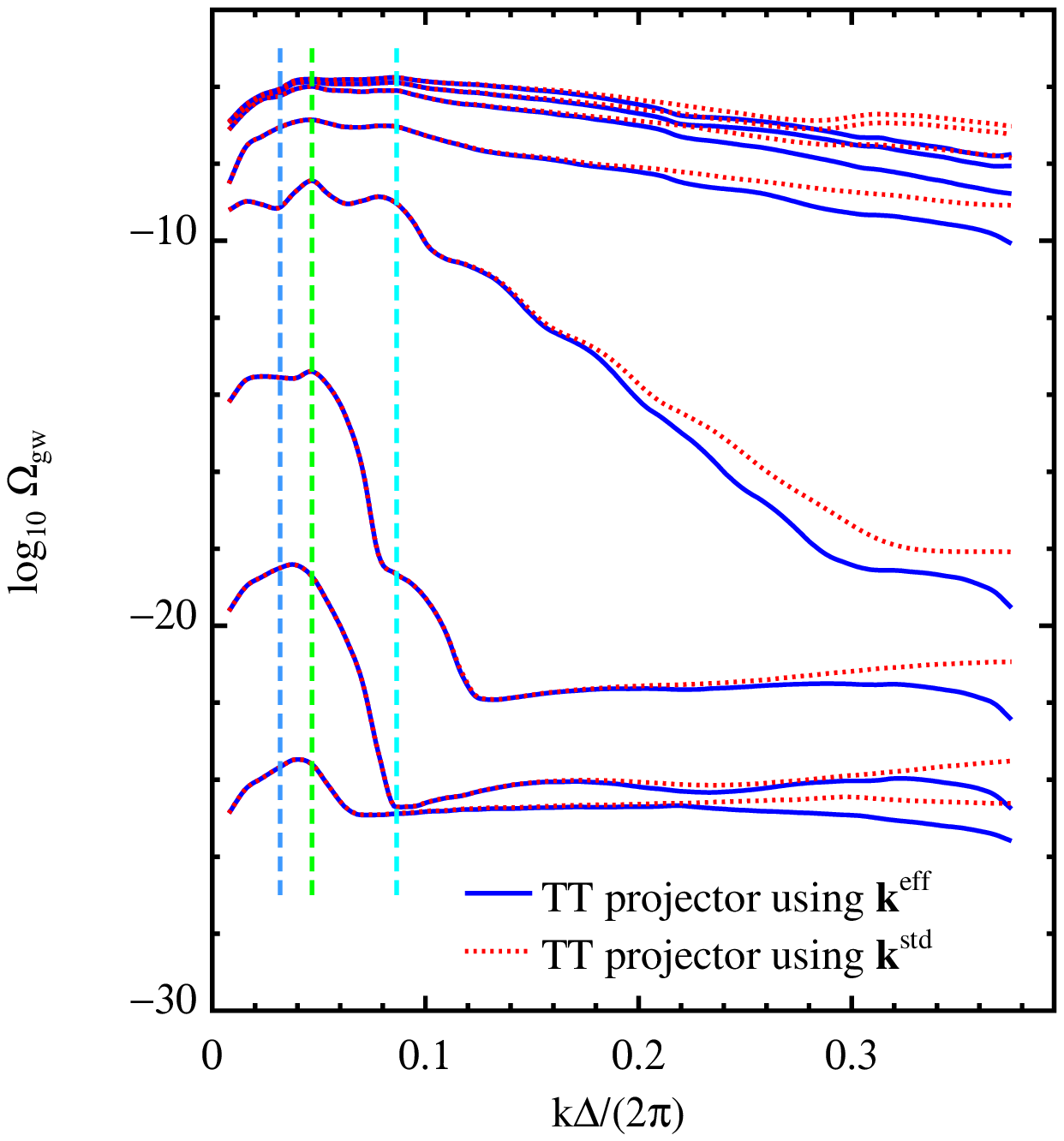}
\includegraphics[width=\figurewidth]{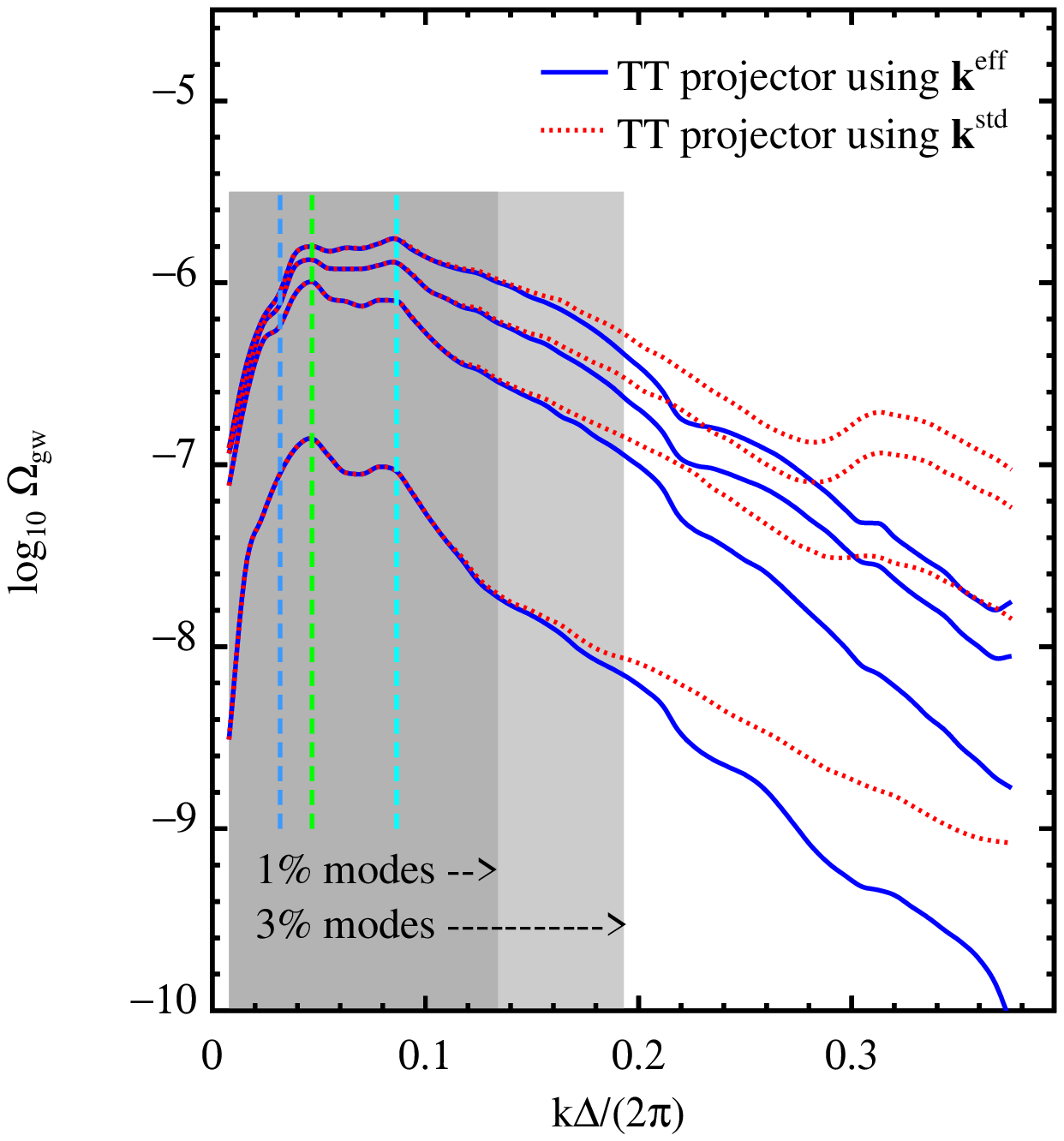}
\caption{The gravitational wave for the model $V(\phi,\chi) = \frac{\lambda}{4}\phi^4+\frac{g^2}{2}\phi^2\chi^2$ with  $\lambda=10^{-14}$ and $g^2/\lambda = 120$. The simulation uses the HLATTICE~V2.0 discretization scheme and ignores metric feedback. The box size is  $L=128\Delta=20/(\sqrt{\lambda}\phi_{\rm ini})$. The blue solid line and red dashed line are GW projected with $\mtt{\vkeff}$ and with $\mtt{\vkstd}$, respectively. In both cases a cutoff $|j_1|, |j_2|, |j_3|<38$ in Fourier space has been used. In the upper panel the output at $a=20$, $30$, ..., $90$ are shown from the bottom to top. In the lower panel only the outputs in the nonlinear regime (from bottom to top, $a=60$, $70$, ..., $90$) are shown. The shaded regions in the lower panel are the regions containing 1\% (dark) and 3\% (dark and light) of the $\vkstd$ modes in the fundamental box ($-n/2<j_1,j_2,j_3\le n/2$). In both panels the characteristic scales shown in \reffig{muk} are plotted with the same color code.\label{gwcomp}}
\end{figure}

The output of GW energy spectra is shown in Figure~\ref{gwcomp}. In the IR part, both the GW projected by $\mtt{\vkeff}$ and that by $\mtt{\vkstd}$ agree with the results found in previous works~\cite{Garcia2008, Easther2007, Dufaux2007}. On the intermediate scales, the GW mapped by $\mtt{\vkstd}$ is significantly higher than that by $\mtt{\vkeff}$, though $\vkeff$ and $\vkstd$ are close to each other. In this example, only $1\%$-$3\%$ of the $128^3$ wave vectors are in the trustable region, the shaded region of the lower panel in Figure~\ref{gwcomp}, where the difference between $\mtt{\vkstd}$ and $\mtt{\vkeff}$ is negligible . This is also $r$ dependent. For a problem with smaller $r$, the trustable region can be even smaller. We have shown in \refsec{subsec:discretization} that at the linear level a mode on the lattice exactly follows the EOM with $k=\keff$ and that $\mtt{\vkeff}$ can completely remove the scalar and vector components defined by the discrete derivatives. Therefore, we may argue that we should trust $\mtt{\vkeff}$ rather than $\mtt{\vkstd}$. However, this argument becomes vague in the nonlinear regime. More discussion along this line is given in \refsec{sec:conclusion}. Nevertheless, what has been explicitly shown here is that caution needs to be taken for most of the modes on the lattice.

\begin{figure}
\includegraphics[width=\figurewidth]{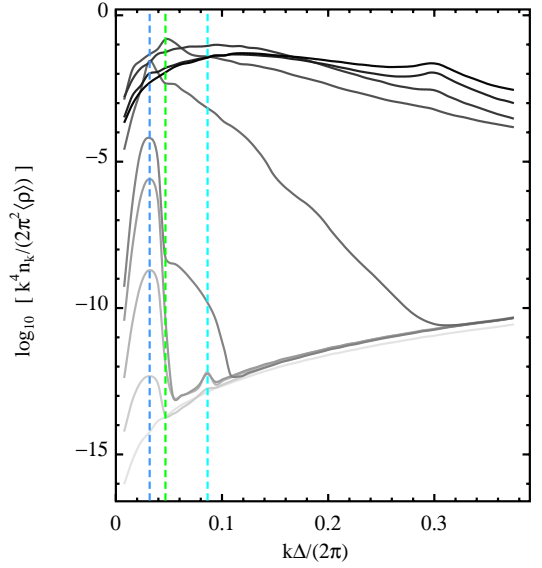}
\caption{The energy power spectra of the $\chi$ field for the preheating model $V(\phi,\chi) = \frac{\lambda}{4}\phi^4+\frac{g^2}{2}\phi^2\chi^2$ with  $\lambda=10^{-14}$ and $g^2/\lambda = 120$. Here $k$ is the effective wavenumber $\keff$; $n_k$ is defined as $n_k\equiv\frac{1}{2k}\left[k^2\abs{\chi_k}^2+\abs{\chi_k'}^2\right]$; $\langle\rho\rangle$ is the mean energy density. The simulation uses the HLATTICE~V2.0 discretization scheme and ignores metric feedback. The box size is  $L=128\Delta=20/(\sqrt{\lambda}\phi_{\rm ini})$. The gray lines from light to dark correspond to $a=1$, $10$, $20$, $30$, ..., $90$. The characteristic scales shown in Figure~\ref{muk} are plotted with the same color code.\label{pwcomp}}
\end{figure}

In \reffig{pwcomp} the energy spectra of the $\chi$ field are plotted, where $n_k$ is defined as $n_k\equiv\frac{1}{2k}\left[k^2\abs{\chi_k}^2+\abs{\chi_k'}^2\right]$. (I have used the wavenumber $k$ instead of the frequency $\omega$ to avoid ambiguity in the nonlinear regime, where $\omega$ is ill-defined.) The growth of $n_k$ in the linear regime ($a\lesssim 30$) agrees excellently with the theoretical expectation shown in \reffig{muk}. In the nonlinear regime, the energy spectrum still peaks around the resonance band until $a\sim 70$. Energy cascading becomes important at $a\gtrsim 70$. When the energy is peaked in a narrow band, energy cascading is efficient. The energy spectrum is soon smoothed at $a\sim 80$. UV cascading slowly goes on after $a\sim 80$. However, due to the finite resolution of the simulation, UV cascading gradually becomes nonphysical on the lattice. When the energy spectrum becomes flat, UV cascading will be strongly affected by the lattice UV cutoff and should no longer be trusted. Physically, UV cascading in a continuum will continue until it is cut off by the quantum effect at very high energy scales where $k^4$ is comparable to the background energy density. In this sense, the classical lattice simulation cannot predict a ``final shape'' of the spectrum. The hope is that, however, other physics such as reheating will take place to stop the cascading at some diffusion scale.

\section{Discussion and Conclusions \label{sec:conclusion}}

In \refsec{subsec:discretization} it was shown that a TT projector $\mtt{\vkstd}$ leads to a numerical noise in GW, which, depending on the tensor-to-scalar ratio and the UV cutoff used in the calculation, may or may not be negligible. This theoretical prediction has been confirmed in \refsec{subsec:chaotic}, where the difference between GW amplitudes calculated with different TT projectors is explicitly shown. 

\begin{figure}
(a1) \includegraphics[width=\halffigurewidth]{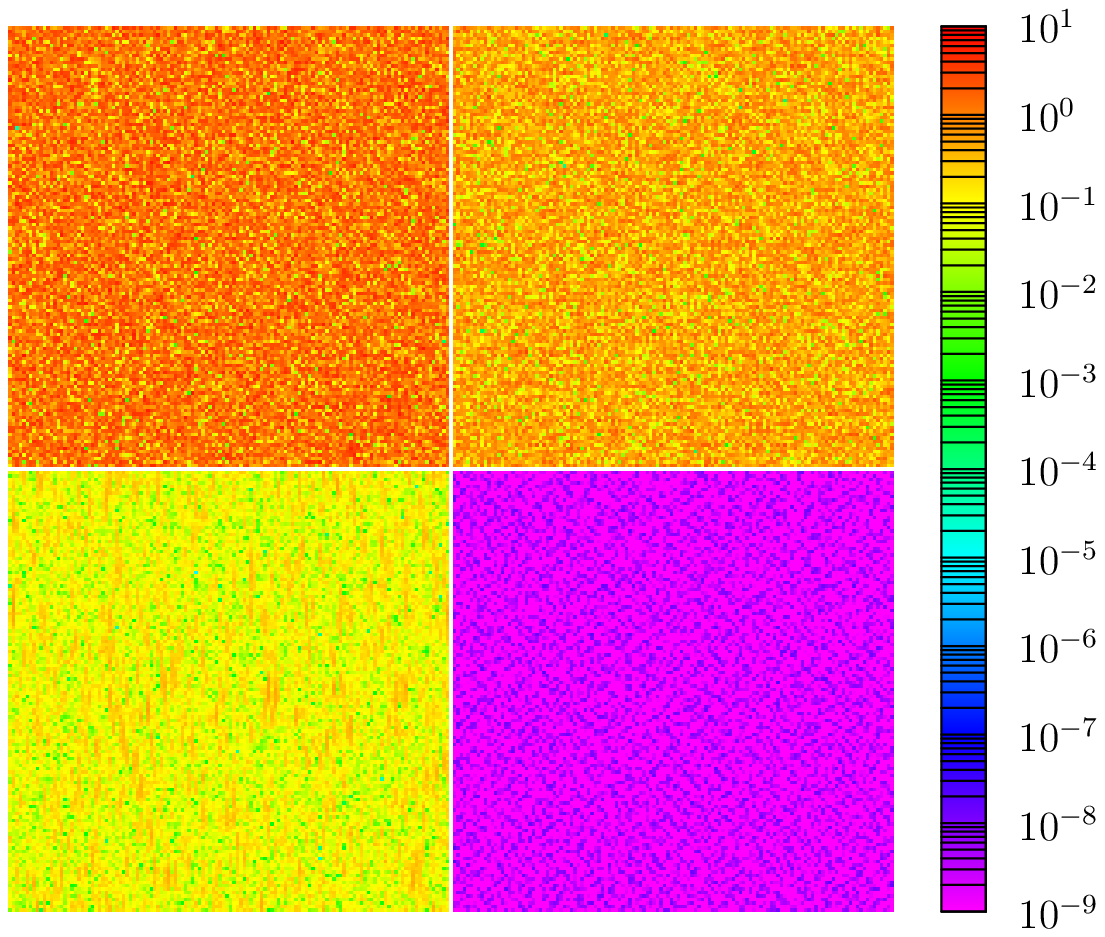}%
(b1) \includegraphics[width=\halffigurewidth]{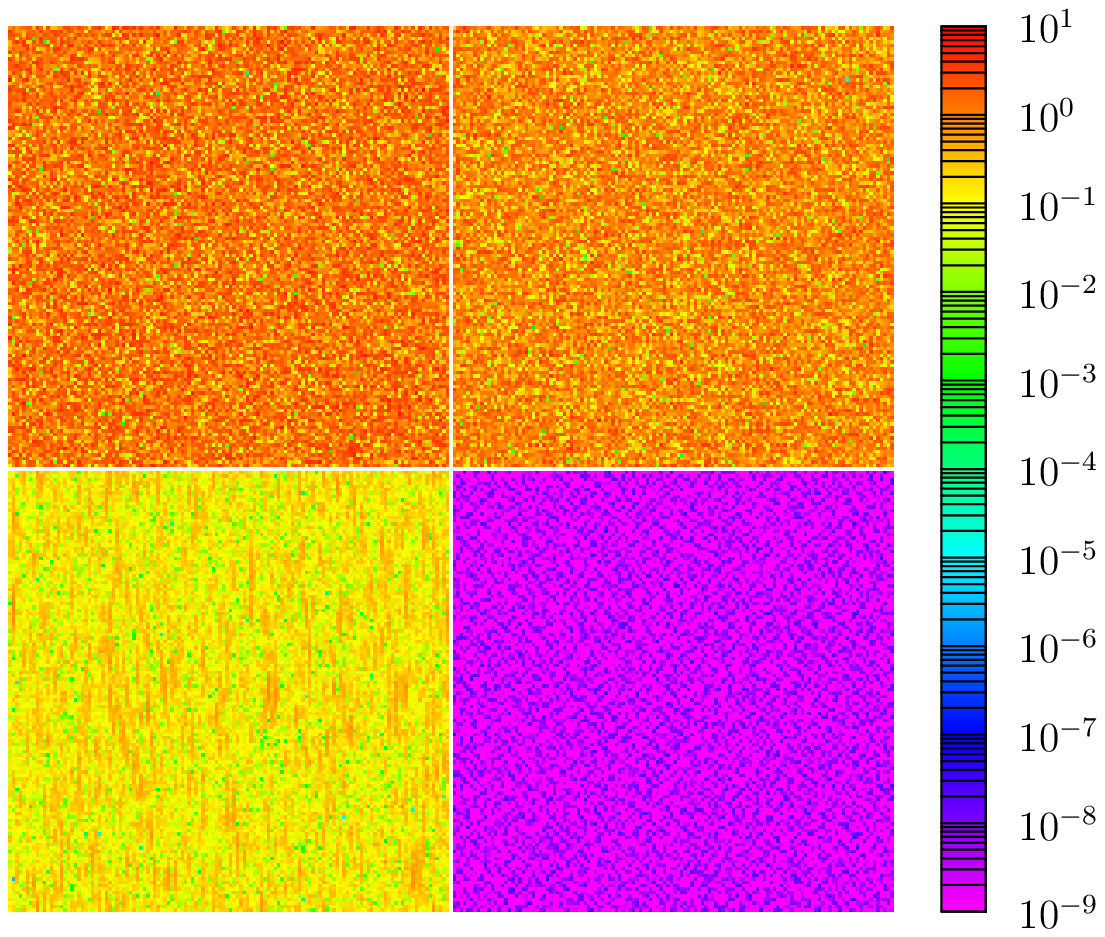} \\
(a2) \includegraphics[width=\halffigurewidth]{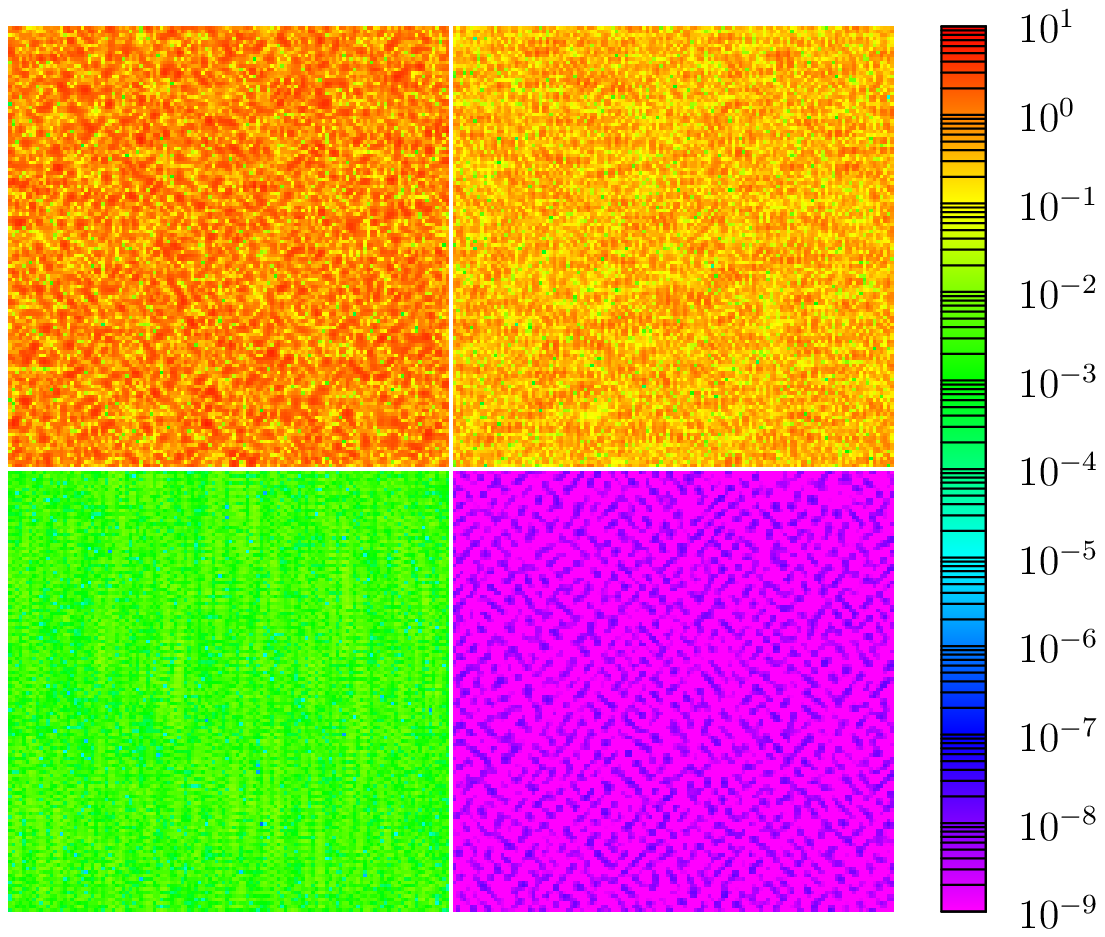}%
(b2) \includegraphics[width=\halffigurewidth]{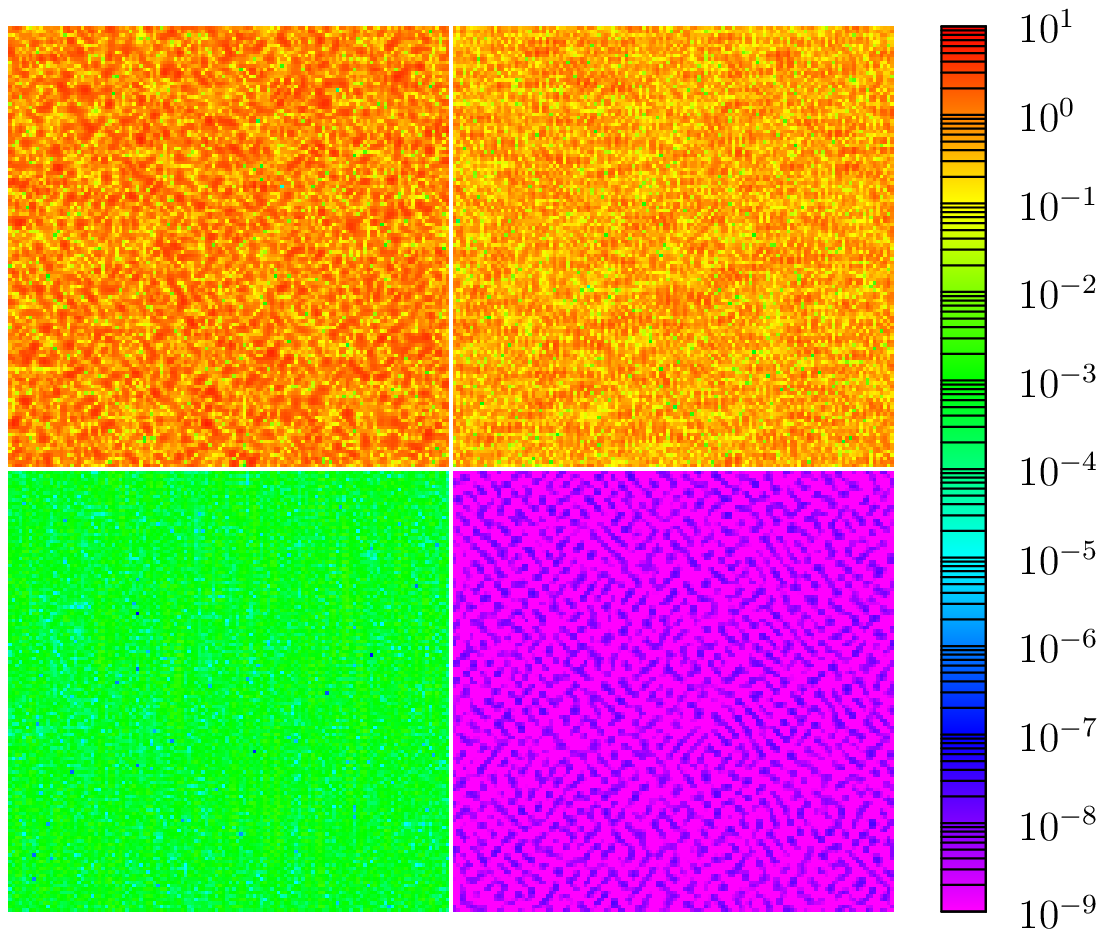}\\
(a3) \includegraphics[width=\halffigurewidth]{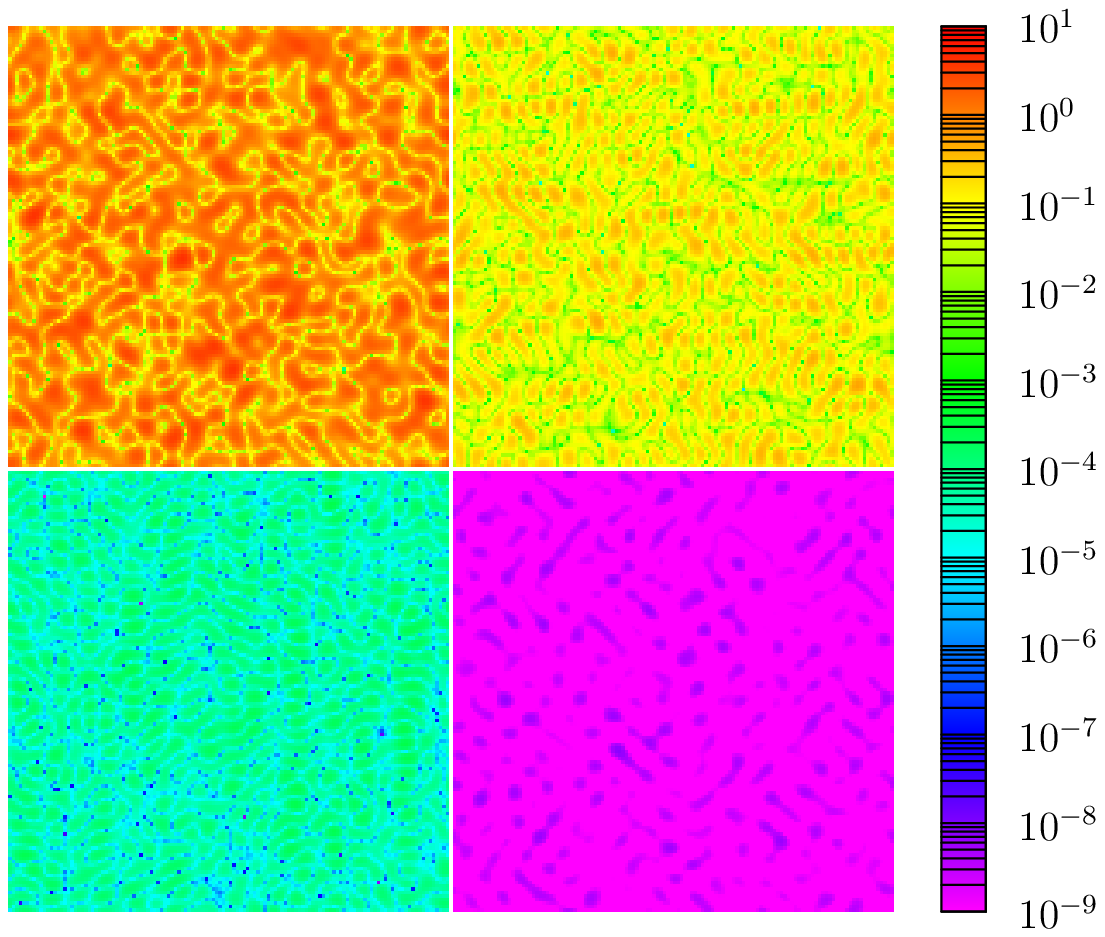}%
(b3) \includegraphics[width=\halffigurewidth]{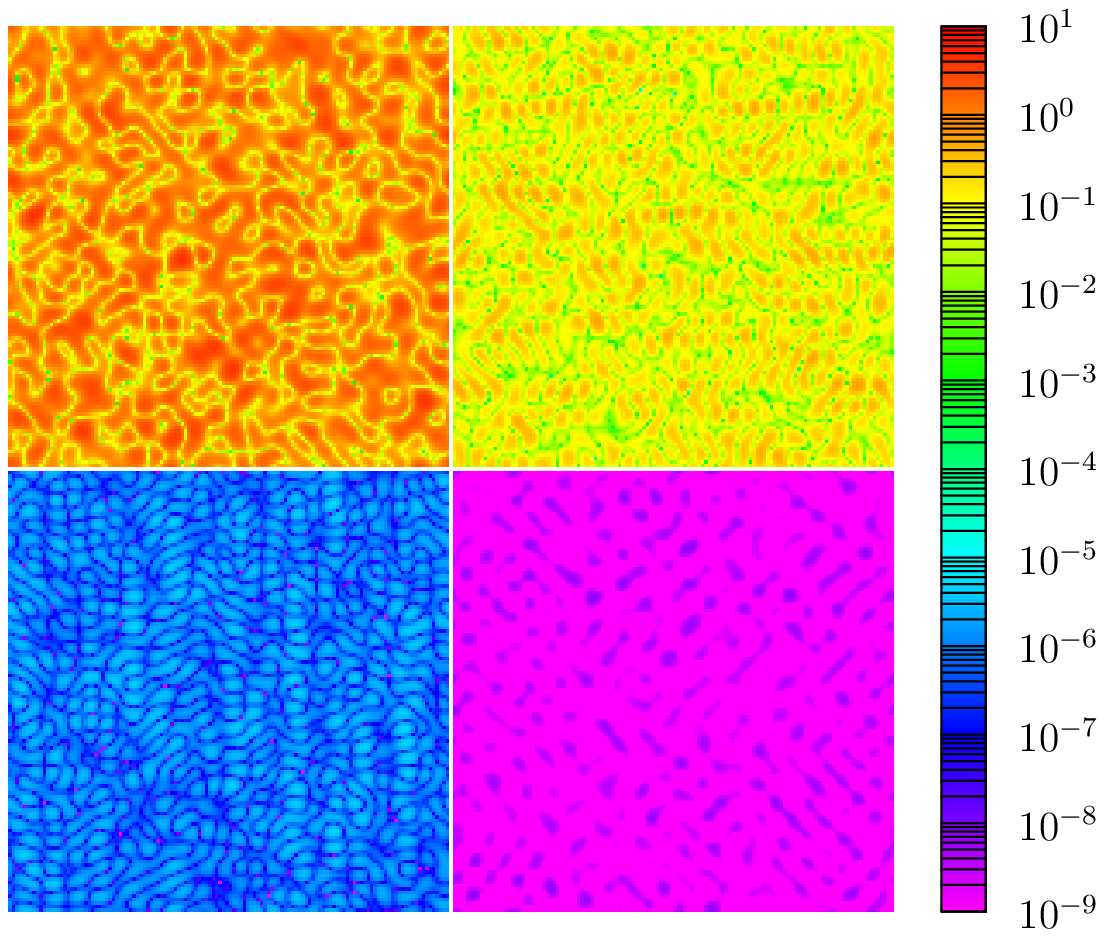}
\caption{Mapping a test field with TT projectors. Each sub-figure contains four panels: the upper-left panel is a 2D slice of the test field $\Lambda$; the upper-right panel is a 2D slice of $|F_{23}|$, where $F_{ij}\equiv \partial_i\partial_j \Lambda$; the lower-left panel is a 2D slice of $|F_{23}|$ after TT projection under $\mtt{\vkstd}$; the lower-right panel is a 2D slice of $|F_{23}|$ after TT projection under $\mtt{\vkeff}$. In sub-figures (a1-a3), the discrete $\partial_i$ is defined in Eqs.~\refeq{derv1}. In sub-figure (b1-b3), the discrete $\partial_i$ is defined in Eqs.~\refeq{derv2}. In sub-figures (a2) and (b2), a Fourier-space cutoff $|j_1|,|j_2|, |j_3|<32$ is applied when generating the $\Lambda$ field, while this cutoff is $|j_1|,|j_2|, |j_3|<16$ for sub-figures (a3) and (b3). See the text for more details. \label{nulltest}}
\end{figure}

It is also shown that the default TT projector in HLATTICE, $\mtt{\vkeff}$, can perfectly remove the scalar and vector components defined by the discrete form of Eq.~\refeq{TTdef_real}. Since defining the discrete derivatives and $\mtt{\vect{k}}$ is a complicated part of the code, it is better to check if the actual code agrees with this theoretical prediction. In \reffig{nulltest} I show a null test. In the null test, a random Gaussian field $\Lambda$ with a scale-invariant spectrum and r.m.s. amplitude $\langle\delta\Lambda^2\rangle^{1/2}=1$ is generated on a lattice with resolution $n=128$. This field is shown in the upper-left panel of each sub-figure. When generating the random field $\Lambda$, a cutoff  $|j_1|,|j_2|, |j_3|<32$ is used in sub-figure (a2) and (b2), and  $|j_1|,|j_2|, |j_3|<16$ in (a3) and (b3).  The discrete derivatives $F_{ij}\equiv \partial_i\partial_j\Lambda$ are calculated using either the HLATTICE~V1.0 discretization scheme (left column, sub-figures a1-a3) or the HLATTICE~V2.0 discretization scheme (right column, sub-figures b1-b3). One component $|F_{23}|$ is shown in the upper-right panel of each sub-figure. The DFT of $F_{ij}$ projected under $\mtt{\vkstd}$ are inverse Fourier transformed back to configuration space. The component $|F_{23}|$ after the TT projection is shown in the lower-left panel of each sub-figure. Similarly, $|F_{23}|$ after TT projection with $\mtt{\vkeff}$ is shown in the lower-right panel of each sub-figure. 

We have confirmed that in the code the scalar component $\partial_i\partial_j\Lambda$ vanishes under the TT projector $\mtt{\vkeff}$. (The $10^{-9}$ level noise is due to the round-off errors in the DFT and inverse DFT calculations.) With the same cutoff in sub-figures (a3) and (b3), the difference between $\mtt{\vkstd}$ and $\mtt{\vkeff}$ is much smaller in HLATTICE~V2.0. This again confirms that $\nabla_{\rm disc}$ defined by Eqs.~\refeq{derv2} is a better approximation to $\nabla|_{\rm cont}$.

Although the projector $\mtt{\vkeff}$ perfectly matches the configuration-space definition of GWs, caution still needs to be taken when we obtain different result using $\mtt{\vkstd}$ and $\mtt{\vkeff}$. The source of GWs, $\partial_i\phi_\ell\partial_j\phi_\ell$, in Fourier space is a convolution,
\begin{eqnarray}
\left.\dft{\partial_i\phi_\ell\partial_j\phi_\ell}\right\vert_{j_1,j_2,j_3} &=& \frac{1}{n^3}\sum \delta^{(3)}(\vkstd_{j'_1,j'_2,j'_3} + \vkstd_{j''_1,j''_2,j''_3}-\vkstd_{j_1,j_2,j_3}) \nonumber \\
&& \times \left.(\ii\keff_i\dft{\phi}_\ell)\right\vert_{j'_1,j'_2,j'_3}\times\left.(\ii\keff_j\dft{\phi}_\ell)\right\vert_{j''_1,j''_2,j''_3},
\end{eqnarray}
where $-n/2<j'_1, j'_2, j'_3, j''_1, j''_2, j''_3\le n/2$. The discrete Kronecker delta  $\delta^{(3)}(\vect{k})$ is $1$ when $\vect{k}=\frac{2\pi}{\Delta}\left(n_1,n_2,n_3\right)$ ($n_1,n_2,n_3=0,\pm 1,\pm 2,...$) and zero otherwise. We see that while the discrete $\nabla$ is mapped to $\ii\vkeff$, the mode-mode coupling is described by $\vkstd$ in the $\delta^{(3)}$ function. The danger of interpreting $\vkeff$ as the physical wave vector is that the mode-mode coupling may be inaccurately described.

An ultimate solution to avoid the ambiguity in the TT projector might be to evolve everything in Fourier space without involving a discrete $\nabla$ approximation. The difficulty, however, is that calculating $\partial V/\partial\phi_\ell$  in Fourier space generally involves convolutions, which is expensive for large $n$. An attempt in this direction is the PSpectRe code by Easther et al. \cite{Easther2010}.

During preheating, the source terms $\partial_i\phi_\ell\partial_j\phi_\ell$ are comparable to the background energy density. In HLATTICE the higher-order gravity self-interaction terms $\lesssim G_g h_{ij}$ in the Lagrangian are ignored. This is valid at least on large scales where the average $G_g$ is much smaller than the background energy density. On smaller scales this might not be a good approximation. In Ref.~\cite{Bastero2010} the authors integrate discretized Einstein equations, and find GW enhanced by an order of magnitude when gravity self-interactions are included. However, the nonlinear enhancement that they find is on large scales (see Figs.~2 and 3 in \cite{Bastero2010}). This is a puzzling result. Note that \cite{Bastero2010} suffers from the same scalar-tensor mixing problem. Also, as discussed in \refsec{subsec:discretization}, the discretization of gravity is not a trivial problem. The authors of \cite{Bastero2010} find that their result is sensitive to the initial conditions of the metric. This is a hint that numerical tachyonic instabilities might exist in their discretization scheme for gravity, because physically a weak gravitational field should not have chaotic feature. Moreover, numerical noises could arise if the integrator is not accurate enough. Ideally their results can be checked by adding all the gravity self-interaction terms into HLATTICE. However, this will significantly complicate the code and  make the simulations much more expensive. I leave this for future work.

In HLATTICE the scalar fields are all assumed to be canonical. An earlier version of HLATTICE, before metric perturbations were incorporated, can simulate noncanonical scalar fields as well. The noncanonical operators are similarly integrated using a Runge-Kutta subintegrator. I will merge the two versions together in the future versions of HLATTICE. The purpose is to accurately study GW produced in preheating with noncanonical scalar fields \cite{Barnaby2009c}.

As a general lattice code with a superior accurate integrator, HLATTICE can be used in many other fields of cosmology. It can be used to study scalar metric perturbations, such as the comoving curvature perturbations studied in \cite{Bond2009}, or nonlinear problems for stochastic inflation models \cite{Salopek1990}. If vector fields can be incorporated, it can also be used to study the electroweak phase transition. To make the code more productive, I release the source code to the community at {\small http://www.cita.utoronto.ca/$\sim$zqhuang/hlat\,}.

\acknowledgements{I thank Andrei Frolov, J.~Richard Bond, Neil Barnaby, Pascal Vaudrevange, and Filippo Vernizzi for useful discussions.}

\end{document}